\documentstyle[preprint,epsfig,eqsecnum,aps]{revtex}
\begin{document}
 \draft
 \title{Phenomenological Analysis of Charmless\\
        Decays $B{\to}PV$ with QCD Factorization}
 \author{Dongsheng Du$^{a,b}$,
 \thanks{Email address:duds@mail.ihep.ac.cn} 
         Haijun Gong$^{b}$,
 \thanks{Email address:gonghj@mail.ihep.ac.cn} 
         Junfeng Sun$^{b}$,
 \thanks{Email address:sunjf@mail.ihep.ac.cn} 
         Deshan Yang$^{b}$
 \thanks{Email address:yangds@mail.ihep.ac.cn} 
         and Guohuai Zhu$^{b,c}$
 \thanks{Email address:zhugh@post.kek.jp}}
 \address{ a. CCAST (World Laboratory),
              P.O.Box 8730,
              Beijing 100080, China \\
           b. Institute of High Energy Physics,
              Chinese Academy of Sciences, \\
              P.O.Box 918(4),
              Beijing 100039, China
 \thanks{Mailing address} \\
           c. Theory Group, KEK, Tsukuba, Ibaraki 305-0801, Japan}
 %\date{}
 \maketitle
 \begin{abstract}
 We study hadronic charmless two-body B decays to final states involving
 pseudoscalar and vector mesons with the QCD factorization approach,
 including the contributions from the chirally enhanced power corrections
 and weak annihilations. The CP-averaged branching ratios and CP-violating
 asymmetries are given. Most of our results are in agreement with the
 present measurements, but the branching ratios of some decay channels are
 only marginally consistent with the experimental observations. 
 Considering the large uncertainties, great advances in both experiment
 and theory in the near future are strongly expected.
 \end{abstract}
 
 \section{introduction}
 \label{sec1}
 In the standard model (SM), CP violation is described by the phase
 of the Cabibbo-Kobayashi-Maskawa (CKM) matrix. Phenomenologically, it
 is clear and convenient to explore CP violation with the well-known
 unitarity triangle
 $V_{ud}V_{ub}^{\ast}+V_{cd}V_{cb}^{\ast}+V_{td}V_{tb}^{\ast}=0$.
 In a sense, the study of $B$ meson decays is mainly to make enough
 independent measurements of the sides and angles of this unitarity
 triangle. It is thus crucial to have a clear understanding of exclusive
 hadronic charmless B decays which would give some useful information
 about and/or constraints on the unitarity triangle. Recently, several
 experimental groups have reported their latest results \cite{0104041,%
 0201007,0115,0111037,0105001,0105019,0107037,0107058,0108017,0109006,%
 0109007,9908018,9912059,0001009,0006008,0101032}, and more B decay
 channels will be measured with great precision soon. With the accumulation
 of the experimental data, theorists are urged to gain deep insight into
 the rare hadronic B decays, and reduce the theoretical uncertainties
 in determining the CKM parameters from experimental measurements.

 Theoretically, intense investigations of hadronic charmless two-body B
 decays have been carried out in great detail with the effective Hamiltonian.
 Combining operator product expansion with the renormalization group
 method, the effective Hamiltonian for B decays is generally expressed by
 the products of the Wilson coefficients and dimension-6 effective 
 operators. The Wilson coefficients can be calculated reliably by 
 perturbation theory, they have been evaluated to the next-to-leading
 order \cite{9512380}. Thus, the main task for us is to evaluate the
 hadronic matrix elements of the effective operators. Based on the naive
 factorization (NF) hypothesis \cite{bsw}, which is supported by the
 argument of color transparency \cite{bjorken}, the hadronic matrix
 elements are usually parametrized into the product of the decay
 constants and the transition form factors phenomenologically. The NF
 approach works well for B and D decays, at least it can give the proper
 orders of magnitude of the branching ratios for B and D decays. However,
 the NF approach is a very rough approximation, and has obvious 
 shortcomings: 
 (1) In the NF framework, the renormalization scheme- and scale-dependence
 in the hadronic matrix elements are completely missing. Its predictions
 would be physical only when the hadronic matrix elements could make
 compensation for the renormalization scheme- and scale-dependence of the
 Wilson coefficients.
 (2) There is no direct CP violation in hadronic B decays with the NF
 approach because the Wilson coefficients, decay constants, and form
 factors are all real. This indicates that ``nonfactorizable'' 
 contributions, which account for final-state rescattering and the strong
 interaction phase shift, are important.
 (3) As stated in \cite{9910291}, it is questionable whether 
 ``nonfactorizable'' effects can be simply absorbed into decay constants
 and form factors. They will have great effects on the class II ($a_{2}$
 dominant) decay modes, e.g. 
 $B^{0}{\to}{\rho}^{0}{\pi}^{0},{\rho}^{0}{\eta},{\omega}{\eta},{\cdots}$.
 So it is imperative to go beyond this level of understanding.

 Recently, M. Beneke, G. Buchalla, M. Neubert, and C. T. Sachrajda
 suggested a QCD factorization (QCDF) method for hadronic B decays in 
 the heavy quark limit, combining the hard-scattering approach with power
 counting in $1/m_{b}$ \cite{QCDF}. The hadronic matrix elements
 ${\langle}M_{1}M_{2}{\vert}O_{i}{\vert}B{\rangle}$ (here $M_{1}$ denotes
 the recoiled meson which picks up the light spectator quark in the B
 meson, $M_{2}$ is the emitted meson, and $O_{i}$ are effective operators)
 can be computed from first principles and expressed in terms of form
 factors and meson light-cone distribution amplitudes (LCDAs) only if
 $M_{2}$ is light or an onium. In addition, when $M_{1}$ is light, there
 exist hard-scattering interactions between $M_{2}$ and the spectator
 which do not exist in the NF approach. The QCDF approach shows that the
 ``nonfactorizable'' contributions are dominated by hard gluon exchange
 and therefore computable systematically. Detailed proofs and arguments
 can be found in \cite{QCDF}; its applications to hadronic two-body B
 decays can be found in the literature \cite{QCDF,0104110,0104090,%
 0012152,0007038,0006022,dyz,0108141}.

 In recent work \cite{0108141}, we calculated the branching ratios and CP
 asymmetries for $B{\to}PP$ (where $P$ denotes the pseudoscalar meson) 
 within the QCDF approach. We found that with appropriate parameters our
 results are in agreement with the current experimental data. This
 encouraged us to further investigate the exclusive decays $B{\to}PV$
 (where $V$ denotes the vector meson) with the QCDF method. The 
 $B{\to}PV$ decays have been carefully studied within generalized
 factorization \cite{9910291,ali,9903453}, and some decay channels have
 also been studied with the perturbative QCD approach \cite{0011238} where
 the form factors are believed to be perturbatively calculable with the
 assistance of the Sudakov factors. However, it is still an open question
 whether the Sudakov factor is applicable in B meson decays 
 \cite{0109260}. In the QCDF framework, it is argued \cite{QCDF} that the
 form factors are not fully calculable according to naive power counting.
 M. Yang and Y. Yang studied the decay modes $B{\to}h_{1}h_{2}$
 ($h_{1}={\pi},K$ and $h_{2}=K^{\ast},{\rho},{\omega}$) in \cite{0007038}
 using the QCDF approach. The differences between their work and ours are
 that we consider the chirally enhanced contributions from another type of
 twist-3 LCDAs $\phi_{\sigma}$ of the pseudoscalar mesons, which is 
 crucial for gauge independence, and we also take the weak annihilation
 topologies into account. H. Cheng and K. Yang pointed out that the
 annihilation contributions could be sizeable for the decays
 $B{\to}{\phi}K$ \cite{0012152}. Sometimes, the annihilation contributions
 may even be dominant, e.g., $B^{0}{\to}K^{\pm}K^{{\ast}{\mp}}$. In this
 paper, we will undertake a comprehensive study of the hadronic charmless
 decays of $B{\to}PV$ within the QCDF framework, including the effects of
 weak annihilation and the chirally enhanced contributions.

 This paper is organized as follows. Section \ref{sec2} is devoted to the
 theoretical framework. There we calculate the ``nonfactorizable''
 corrections to hadronic matrix elements of the effective operators with
 the QCDF approach, including the chirally enhanced power corrections and
 weak annihilations. The input parameters in our calculations are given in
 Sec. \ref{sec3}. The numerical values and some remarks about the
 CP-averaged branching ratios and CP-violating asymmetries for $B{\to}PV$
 decays are arranged in Sec. \ref{sec4} and Sec. \ref{sec5}, respectively.
 We come to our final conclusions in Sec. \ref{sec6}.

 \section{Theoretical framework for B decays}
 \label{sec2}

 \subsection{The effective Hamiltonian}
 \label{sec21}
 The effective Hamiltonian for hadronic charmless B decays is written
 as \cite{9512380}:
 \begin{eqnarray}
 {\cal H}_{eff}&=&\frac{G_{F}}{\sqrt{2}}
 \Big\{ {\sum\limits_{q=u,c}} v_{q}
     \Big[ C_{1}({\mu}) Q^{q}_{1}({\mu})
         + C_{2}({\mu}) Q^{q}_{2}({\mu})
         +{\sum\limits_{k=3}^{10}} C_{k}({\mu}) Q_{k}({\mu})
     \Big] \nonumber \\
 & & -v_{t} \Big[ C_{7{\gamma}} Q_{7{\gamma}}
                + C_{8g} Q_{8g} \Big] \Big\} + H.c. ,
 \label{eq:Hamiltonian}
 \end{eqnarray}
 where $v_{q}=V_{qb}V_{qd}^{\ast}$ (for $b{\to}d$ transitions) or
       $v_{q}=V_{qb}V_{qs}^{\ast}$ (for $b{\to}s$ transitions) are CKM
 factors. $C_{i}({\mu})$ are Wilson coefficients; they are universal and
 process independent, and have been evaluated to the next-to-leading
 logarithmic order (NLO) with the perturbation theory and renormalization
 group method. We list their numerical values in the naive dimensional
 regularization (NDR) scheme at three different scales in Table \ref{tab1}.
 The dimension-6 local operators, including tree operators $Q_{1}^{q}{\sim}
 Q_{2}^{q}$, QCD penguin operators $Q_{3}{\sim}Q_{6}$, electroweak
 penguin operators $Q_{7}{\sim}Q_{10}$, and magnetic penguin operators
 $Q_{7{\gamma}},Q_{8g}$, can be expressed explicitly as
 \begin{mathletters}
 \begin{eqnarray}
  & &Q^{u}_{1}=({\bar{u}}_{\alpha}b_{\alpha})_{V-A}
               ({\bar{q}}_{\beta} u_{\beta} )_{V-A},
     \ \ \ \ \ \ \ \ \ \ \ \ \ \ \ \ \ \ \
     Q^{c}_{1}=({\bar{c}}_{\alpha}b_{\alpha})_{V-A}
               ({\bar{q}}_{\beta} c_{\beta} )_{V-A}, \\
  & &Q^{u}_{2}=({\bar{u}}_{\alpha}b_{\beta} )_{V-A}
               ({\bar{q}}_{\beta} u_{\alpha})_{V-A},
     \ \ \ \ \ \ \ \ \ \ \ \ \ \ \ \ \ \ \
     Q^{c}_{2}=({\bar{c}}_{\alpha}b_{\beta} )_{V-A}
            ({\bar{q}}_{\beta} c_{\alpha})_{V-A}, \\
  & &Q_{3}=({\bar{q}}_{\alpha}b_{\alpha})_{V-A}\sum\limits_{q^{\prime}}
           ({\bar{q}}^{\prime}_{\beta} q^{\prime}_{\beta} )_{V-A},
     \ \ \ \ \ \ \ \ \ \ \ \ \ \ \ \
     Q_{4}=({\bar{q}}_{\beta} b_{\alpha})_{V-A}\sum\limits_{q^{\prime}}
           ({\bar{q}}^{\prime}_{\alpha}q^{\prime}_{\beta} )_{V-A}, \\
  & &Q_{5}=({\bar{q}}_{\alpha}b_{\alpha})_{V-A}\sum\limits_{q^{\prime}}
           ({\bar{q}}^{\prime}_{\beta} q^{\prime}_{\beta} )_{V+A},
     \ \ \ \ \ \ \ \ \ \ \ \ \ \ \ \
     Q_{6}=({\bar{q}}_{\beta} b_{\alpha})_{V-A}\sum\limits_{q^{\prime}}
           ({\bar{q}}^{\prime}_{\alpha}q^{\prime}_{\beta} )_{V+A}, \\
  & &Q_{7}=\frac{3}{2}({\bar{q}}_{\alpha}b_{\alpha})_{V-A}
           \sum\limits_{q^{\prime}}e_{q^{\prime}}
           ({\bar{q}}^{\prime}_{\beta} q^{\prime}_{\beta} )_{V+A},
     \ \ \ \ \ \ \ \ \ \
     Q_{8}=\frac{3}{2}({\bar{q}}_{\beta} b_{\alpha})_{V-A}
           \sum\limits_{q^{\prime}}e_{q^{\prime}}
           ({\bar{q}}^{\prime}_{\alpha}q^{\prime}_{\beta} )_{V+A}, \\
  & &Q_{9}=\frac{3}{2}({\bar{q}}_{\alpha}b_{\alpha})_{V-A}
           \sum\limits_{q^{\prime}}e_{q^{\prime}}
           ({\bar{q}}^{\prime}_{\beta} q^{\prime}_{\beta} )_{V-A},
     \ \ \ \ \ \ \ \ \ \
    Q_{10}=\frac{3}{2}({\bar{q}}_{\beta} b_{\alpha})_{V-A}
           \sum\limits_{q^{\prime}}e_{q^{\prime}}
           ({\bar{q}}^{\prime}_{\alpha}q^{\prime}_{\beta} )_{V-A}, \\
  & &Q_{7{\gamma}}=\frac{e}{8{\pi}^{2}}m_{b}{\bar{q}}_{\alpha}
           {\sigma}^{{\mu}{\nu}}(1+{\gamma}_{5})
            b_{\alpha}F_{{\mu}{\nu}},
     \ \ \ \ \ \ \ \ \ \ \
     Q_{8g}=\frac{g}{8{\pi}^{2}}m_{b}{\bar{q}}_{\alpha}
           {\sigma}^{{\mu}{\nu}}(1+{\gamma}_{5})
            t^{a}_{{\alpha}{\beta}}b_{\beta}G^{a}_{{\mu}{\nu}},
 \end{eqnarray}
 \label{eq:operators}
 \end{mathletters}
 where $q^{\prime}$ denotes all the active quarks at the scale
 ${\mu}={\cal O}(m_{b})$, i.e., $q^{\prime}=u,d,s,c,b$.

 \subsection{$B{\to}PV$ in the QCDF framework}
 \label{sec22}
 When the QCDF method is applied to the decays $B{\to}PV$, the hadronic matrix
 elements of the local effective operators can be written as
 \begin{eqnarray}
 {\langle}PV{\vert}O_{i}{\vert}B{\rangle}&=&
    F^{B{\to}P}_{j}(0) {\int}_{0}^{1}dx \ T^{I}_{ij}(x) {\Phi}_{V}(x)
  + A^{B{\to}V}_{k}(0) {\int}_{0}^{1}dy \ T^{I}_{ik}(y) {\Phi}_{P}(y)
    \nonumber \\ & &
  + {\int}_{0}^{1} d{\xi} {\int}_{0}^{1} dx {\int}_{0}^{1} dy \
    T^{II}_{i}({\xi},x,y) {\Phi}_{B}(\xi) {\Phi}_{V}(x) {\Phi}_{P}(y).
 \label{eq:qcdf}
 \end{eqnarray}
 Here $F^{B{\to}P}$ and $A^{B{\to}V}$ denote the form factors for $B{\to}P$
 and $B{\to}V$ transitions, respectively. ${\Phi}_{B}(\xi)$,
 ${\Phi}_{V}(x)$, and ${\Phi}_{P}(y)$ are the LCDAs of valence quark Fock
 states for B, vector, and pseudoscalar mesons, respectively.
 $T^{I,II}_{i}$ denote the hard-scattering kernels, which are dominated by
 hard gluon exchange when the power suppressed 
 ${\cal O}({\Lambda}_{QCD}/m_{b})$ terms are neglected. So they are
 calculable order by order in perturbation theory. The leading terms of
 $T^{I}$ come from the tree level and correspond to the NF approximation.
 The order of ${\alpha}_{s}$ terms of $T^I$ can be depicted by 
 vertex-correction diagrams Fig.1(a-d) and penguin-correction diagrams
 Fig.1(e-f). $T^{II}$ describes the hard interactions between the
 spectator quark and the emitted meson $M_{2}$ when the gluon virtuality
 is large. Its lowest order terms are ${\cal O}({\alpha}_{s})$ and can be
 depicted by hard spectator scattering diagrams Fig.1(g-h). One of the
 most interesting results of the QCDF approach is that, in the heavy quark
 limit, the strong phases arise naturally from the hard-scattering kernels
 at the order of ${\alpha}_{s}$. As for the nonperturbative part, it is
 either power suppressed in $1/m_b$ or separated into the form factors
 and LCDAs of mesons.

 Because the b quark mass is not asymptotically large, the power
 suppression might fail in some cases. For instance, the contributions of
 operator $Q_6$ to the decay amplitudes would formally vanish in the
 strict heavy quark limit. However, it is numerically very important in
 penguin-dominated B rare decays, such as the interesting channels $B{\to}
 {\pi}K$, etc. This is because $Q_6$ is always multiplied by a formally
 power suppressed but chirally enhanced factor 
 $r_{\chi}=\frac{2m_{P}^2}{m_b(m_1+m_2)} \sim {\cal O}(1)$, where $m_1$
 and $m_2$ are current quark masses. Therefore phenomenological 
 application of QCD factorization in B rare decays requires at least a
 consistent inclusion of chirally enhanced corrections. The readers may
 refer to Refs. \cite{0104110,dyz} for more details.

 With the above discussions on the effective Hamiltonian of B decays
 Eq.(\ref{eq:Hamiltonian}) and the QCDF expressions of hadronic matrix
 elements Eq.(\ref{eq:qcdf}), the decay amplitudes for $B{\to}PV$ in the
 heavy quark limit can be written as
 \begin{equation}
 {\cal A}(B{\to}PV) = \frac{G_{F}}{\sqrt{2}}
    \sum\limits_{p=u,c} \sum\limits_{i=1}^{10} v_{p} a_{i}^{p}
   {\langle}PV{\vert}O_{i}{\vert}B{\rangle}_{f}.
 \label{eq:decay-f}
 \end{equation}
 The above ${\langle}PV{\vert}O_{i}{\vert}B{\rangle}_{f}$ are the 
 factorized hadronic matrix elements, which have the same definitions as
 those in the NF approach. The ``nonfactorizable'' effects are included in
 the coefficients $a_{i}$ which are process dependent. The coefficients
 $a_{i}$ are collected in Sec. \ref{sec23}, and the explicit expressions
 for the decay amplitudes of $B{\to}PV$ can be found in Appendix.B of
 \cite{ali}. The only differences from ours are the expressions for the
 coefficients $a_{i}$.

 According to the arguments in \cite{QCDF}, the contributions of weak
 annihilation to the decay amplitudes are power suppressed, and they
 do not appear in the QCDF formula Eq.(\ref{eq:qcdf}). But, as emphasized
 in \cite{0004173,0004213}, the contributions from weak annihilation
 could give large strong phases with QCD corrections, and hence large CP
 violation could be expected, so their effects cannot simply be neglected.
 However, in the QCDF method, the annihilation topologies (see Fig.2)
 violate factorization because of the endpoint divergence. There is
 similar endpoint divergence when considering the chirally enhanced hard
 spectator scattering. One possible way is to treat the endpoint
 divergence from different sources as different phenomenological
 parameters \cite{0104110}. The corresponding price is the introduction 
 of model dependence and extra numerical uncertainties in the decay
 amplitudes. In this work, we will follow the treatment of Ref.
 \cite{0104110} and express the weak annihilation topological decay
 amplitudes as
 \begin{equation}
 {\cal A}^{a}(B{\to}PV) {\propto} f_{B} f_{P} f_{V}
    {\sum} v_{p} b_{i} \; ,
 \label{eq:decay-a}
 \end{equation}
 where the parameters $b_{i}$ are collected in Sec. \ref{sec24}, and the
 expressions for the weak annihilation decay amplitudes of $B{\to}PV$ are
 listed in the Appendix \ref{sec:app1}. To distinguish the decay
 amplitudes Eq.(\ref{eq:decay-f}) from Eq.(\ref{eq:decay-a}), we will add
 a superscript {\em f} to the symbol in Eq.(\ref{eq:decay-f}), and write
 it as ${\cal A}(B{\to}PV){\to}{\cal A}^{f}(B{\to}PV)$.
  
 \subsection{The QCD coefficients $a_{i}$}
 \label{sec23}
 We now present the QCD coefficients $a_{i}$ in Eq.(\ref{eq:decay-f}). In
 our calculations, we neglect the mass of light quarks when we apply the
 equation of motion to the external quarks. We consider the chirally
 enhanced contributions from twist-3 LCDAs of the pseudoscalar mesons. As
 for the vector mesons, only the leading twist LCDAs for the
 longitudinally polarized components ${\Phi}_{V}^{\|}$ contribute to the
 hadronic matrix elements, while the effects of twist-2 LCDAs for the
 transversely polarized components  ${\Phi}_{V}^{\bot}$ and higher twist
 LCDAs of the vector mesons are power suppressed and can therefore be
 neglected within the QCDF framework. We express the coefficients $a_{i}$
 in two parts, i.e., $a_{i}=a_{i,I}+a_{i,II}$. The first term $a_{i,I}$
 contains the ``nonfactorizable'' effects which are described by 
 Fig.1(a-f), while the second part $a_{i,II}$ corresponds to the hard
 spectator scattering diagrams Fig.1(g-h).

 There are two different cases according to the final states. Case I
 is that the recoiled meson $M_{1}$ is a vector meson, and the emitted
 meson $M_{2}$ corresponds to a pseudoscalar meson, and vice versa for
 case II. For case I, we sum up the results for $a_{i}$ as follows:
 \begin{mathletters}
 \begin{eqnarray}
   & &a_{1,I} = C_{1} + \frac{C_{2}}{N_{c}}
                \Big[
                1 + \frac{C_{F}{\alpha}_{s}}{4{\pi}} V_{M}
                \Big],
      \ \ \ \ \ \ \ \ \ \ \ \ \ \ \ \ \ \ \ \ \ \ \ \
      a_{1,II} = \frac{{\pi}C_{F}{\alpha}_{s}}{N_{c}^{2}}
                 C_{2} H(BM_{1},M_{2}), \\
   & &a_{2,I} = C_{2} + \frac{C_{1}}{N_{c}}
                \Big[
                1 + \frac{C_{F}{\alpha}_{s}}{4{\pi}} V_{M}
                \Big],
      \ \ \ \ \ \ \ \ \ \ \ \ \ \ \ \ \ \ \ \ \ \ \ \
      a_{2,II} = \frac{{\pi}C_{F}{\alpha}_{s}}{N_{c}^{2}}
                 C_{1} H(BM_{1},M_{2}), \\
   & &a_{3,I} = C_{3} + \frac{C_{4}}{N_{c}}
                \Big[
                1 + \frac{C_{F}{\alpha}_{s}}{4{\pi}} V_{M}
                \Big],
      \ \ \ \ \ \ \ \ \ \ \ \ \ \ \ \ \ \ \ \ \ \ \ \
      a_{3,II} = \frac{{\pi}C_{F}{\alpha}_{s}}{N_{c}^{2}}
                 C_{4} H(BM_{1},M_{2}), \\
   & &a_{4,I}^{p} = C_{4} + \frac{C_{3}}{N_{c}}
                \Big[
                1 + \frac{C_{F}{\alpha}_{s}}{4{\pi}} V_{M}
                \Big]
                  + \frac{C_{F}{\alpha}_{s}}{4{\pi}}
                    \frac{P_{M,2}^{p}}{N_{c}},
      \ \ \ \ \ \
      a_{4,II} = \frac{{\pi}C_{F}{\alpha}_{s}}{N_{c}^{2}}
                 C_{3} H(BM_{1},M_{2}), \\
   & &a_{5,I} = C_{5} + \frac{C_{6}}{N_{c}}
                \Big[
                1 - \frac{C_{F}{\alpha}_{s}}{4{\pi}} V_{M}^{\prime}
                \Big],
      \ \ \ \ \ \ \ \ \ \ \ \ \ \ \ \ \ \ \ \ \ \ \ \
      a_{5,II} = - \frac{{\pi}C_{F}{\alpha}_{s}}{N_{c}^{2}}
                 C_{6} H^{\prime}(BM_{1},M_{2}), \\
   & &a_{6,I}^{p} = C_{6} + \frac{C_{5}}{N_{c}}
                \Big[
                1 - 6 \frac{C_{F}{\alpha}_{s}}{4{\pi}}
                \Big]
                    + \frac{C_{F}{\alpha}_{s}}{4{\pi}}
                      \frac{P_{M,3}^{p}}{N_{c}},
      \ \ \ \ \ \ \ \ \
      a_{6,II} = 0, \\
   & &a_{7,I} = C_{7} + \frac{C_{8}}{N_{c}}
                \Big[
                1 - \frac{C_{F}{\alpha}_{s}}{4{\pi}} V_{M}^{\prime}
                \Big],
      \ \ \ \ \ \ \ \ \ \ \ \ \ \ \ \ \ \ \ \ \ \ \ \
      a_{7,II} = - \frac{{\pi}C_{F}{\alpha}_{s}}{N_{c}^{2}}
                 C_{8} H^{\prime}(BM_{1},M_{2}), \\
   & &a_{8,I}^{p} = C_{8} + \frac{C_{7}}{N_{c}}
                \Big[
                1 - 6 \frac{C_{F}{\alpha}_{s}}{4{\pi}}
                \Big]
                    + \frac{\alpha}{9{\pi}}
                      \frac{P_{M,3}^{p,ew}}{N_{c}},
      \ \ \ \ \ \ \ \ \ \ \ \
      a_{8,II} = 0, \\
   & &a_{9,I} = C_{9} + \frac{C_{10}}{N_{c}}
                \Big[
                1 + \frac{C_{F}{\alpha}_{s}}{4{\pi}} V_{M}
                \Big],
      \ \ \ \ \ \ \ \ \ \ \ \ \ \ \ \ \ \ \ \ \ \ \ \
      a_{9,II} = \frac{{\pi}C_{F}{\alpha}_{s}}{N_{c}^{2}}
                 C_{10} H(BM_{1},M_{2}), \\
   & &a_{10,I}^{p} = C_{10} + \frac{C_{9}}{N_{c}}
                \Big[
                 1 + \frac{C_{F}{\alpha}_{s}}{4{\pi}} V_{M}
                \Big]
                    + \frac{\alpha}{9{\pi}}
                      \frac{P_{M,2}^{p,ew}}{N_{c}},
      \ \ \ \ \ \ \ \
      a_{10,II} = \frac{{\pi}C_{F}{\alpha}_{s}}{N_{c}^{2}}
                  C_{9} H(BM_{1},M_{2}),
 \end{eqnarray}
 \label{eq:ai}
 \end{mathletters}
 where $C_{F}=\frac{N_{c}^{2}-1}{2N_{c}}$, and $N_{c}=3$. The vertex
 parameters $V_{M}$ and $V_{M}^{\prime}$ result from Fig.1(a-d); the QCD
 penguin parameters $P_{M,i}^{p}$ and the electroweak penguin parameters
 $P_{M,i}^{p,ew}$ result from Fig.1(e-f). The expressions for the penguin
 parameters are a little different from those in \cite{0104110}. Here we
 consider the corrections of the electroweak penguins to $P_{M,i}^{p}$,
 and the contributions of the QCD penguins to $P_{M,i}^{p,ew}$. They can
 be written as
 \begin{mathletters}
 \begin{eqnarray}
  P_{M,2}^{p} &=& C_{1} \Big[ \frac{4}{3} {\ln}\frac{m_{b}}{\mu}
                            + \frac{2}{3} - G_{M}(s_{p}) \Big]
                 + \Big( C_{3} - \frac{1}{2} C_{9} \Big)
                   \Big[ \frac{8}{3} {\ln}\frac{m_{b}}{\mu}
                       + \frac{4}{3} - G_{M}(0) - G_{M}(1) \Big]
             \nonumber \\
             & & + \sum\limits_{q=q^{\prime}} \Big( C_{4} + C_{6}
                       + \frac{3}{2} e_{q} C_{8}
                       + \frac{3}{2} e_{q} C_{10} \Big)
                   \Big[ \frac{4}{3} {\ln}\frac{m_{b}}{\mu}
                       - G_{M}(s_{q}) \Big]
             \nonumber \\
             & & - 2 C_{8g} {\int}_{0}^{1}dx \
                         \frac{{\Phi}_{M}(x)}{1-x}, \\
  P_{M,3}^{p} &=& C_{1} \Big[ \frac{4}{3} {\ln}\frac{m_{b}}{\mu}
                            + \frac{2}{3} - {\hat{G}}_{M}(s_{p}) \Big]
                 + \Big( C_{3} - \frac{1}{2} C_{9} \Big)
                   \Big[ \frac{8}{3} {\ln}\frac{m_{b}}{\mu}
                       + \frac{4}{3} - {\hat{G}}_{M}(0)
                       - {\hat{G}}_{M}(1) \Big]
             \nonumber \\
             & & + \sum\limits_{q=q^{\prime}} \Big( C_{4} + C_{6}
                       + \frac{3}{2} e_{q} C_{8}
                       + \frac{3}{2} e_{q} C_{10} \Big)
                   \Big[ \frac{4}{3} {\ln}\frac{m_{b}}{\mu}
                       - {\hat{G}}_{M}(s_{q}) \Big]
                       - 2 C_{8g}, \\
  P_{M,2}^{p,ew} &=& \Big( C_{1} + N_{c} C_{2} \Big)
                     \Big[ \frac{4}{3} {\ln}\frac{m_{b}}{\mu}
                         + \frac{2}{3} - G_{M}(s_{p}) \Big]
               \nonumber \\
               & & - \Big( C_{3} + N_{c} C_{4} \Big)
                     \Big[ \frac{4}{3} {\ln}\frac{m_{b}}{\mu}
                         + \frac{2}{3} - \frac{1}{2} G_{M}(0)
                         - \frac{1}{2} G_{M}(1) \Big]
               \nonumber \\
               & & + \sum\limits_{q=q^{\prime}} \Big( N_{c} C_{3} + C_{4}
                   + N_{c} C_{5} + C_{6} \Big) \frac{3}{2} e_{q}
                     \Big[ \frac{4}{3} {\ln}\frac{m_{b}}{\mu}
                         - G_{M}(s_{q}) \Big]
               \nonumber \\
               & & - N_{c} C_{7{\gamma}} {\int}_{0}^{1}dx \
                     \frac{{\Phi}_{M}(x)}{1-x}, \\
  P_{M,3}^{p,ew} &=& \Big( C_{1} + N_{c} C_{2} \Big)
                     \Big[ \frac{4}{3} {\ln}\frac{m_{b}}{\mu}
                         + \frac{2}{3} - {\hat{G}}_{M}(s_{p}) \Big]
               \nonumber \\
               & & - \Big( C_{3} + N_{c} C_{4} \Big)
                     \Big[ \frac{4}{3} {\ln}\frac{m_{b}}{\mu}
                         + \frac{2}{3} - \frac{1}{2} {\hat{G}}_{M}(0)
                          - \frac{1}{2} {\hat{G}}_{M}(1) \Big]
               \nonumber \\
               & & + \sum\limits_{q=q^{\prime}} \Big( N_{c} C_{3} + C_{4}
                   + N_{c} C_{5} + C_{6} \Big) \frac{3}{2} e_{q}
                     \Big[ \frac{4}{3} {\ln}\frac{m_{b}}{\mu}
                         - {\hat{G}}_{M}(s_{q}) \Big]
                   - N_{c} C_{7{\gamma}}.
 \end{eqnarray}
 \label{eq:penguin}
 \end{mathletters}
 The definitions of quantities $V_{M}$, $V_{M}^{\prime}$, $G_{M}(s_{q})$,
 and ${\hat{G}}_{M}(s_{q})$ can be found in \cite{0104110}, and
 $s_{q}=m_{q}^{2}/m_{b}^{2}$. In the expressions for $P_{M,i}^{p}$ and
 $P_{M,i}^{p,ew}$, $q^{\prime}$ will run over all the active quarks at
 the  scale ${\mu}={\cal O}(m_{b})$, i.e., $q^{\prime}=u,d,s,c,b$. The
 parameters $H(BM_{1},M_{2})$ and $H^{\prime}(BM_{1},M_{2})$ in 
 $a_{i,II}$, which originate from hard gluon exchanges between the
 spectator quark and the emitted meson $M_{2}$, are written as
 \begin{mathletters}
 \begin{eqnarray}
  H(BV,P)&=&\frac{f_{B}f_{V}}{m_{B}^{2}A_{0}^{B{\to}V}}
  {\int}_{0}^{1} d{\xi} {\int}_{0}^{1} dx {\int}_{0}^{1} dy
   \frac{{\Phi}_{B}({\xi})}{\xi}
   \frac{{\Phi}_{P}(x)}{\bar{x}}
   \frac{{\Phi}_{V}(y)}{\bar{y}},
   \\
  H^{\prime}(BV,P)&=&\frac{f_{B}f_{V}}{m_{B}^{2}A_{0}^{B{\to}V}}
  {\int}_{0}^{1} d{\xi} {\int}_{0}^{1} dx {\int}_{0}^{1} dy
   \frac{{\Phi}_{B}({\xi})}{\xi}
   \frac{{\Phi}_{P}(x)}{x}
   \frac{{\Phi}_{V}(y)}{\bar{y}}.
 \end{eqnarray}
 \label{eq:hard-spectator-p}
 \end{mathletters}

 For case II, except for the parameters of $H(BM_{1},M_{2})$ and
 $H^{\prime}(BM_{1},M_{2})$, the expressions for $a_{i}$ are similar to
 those in case I. In particular, we would like to point out that because
 ${\langle}V{\vert}(\bar{q}q)_{S{\pm}P}{\vert}0{\rangle}=0$, the
 contributions of the effective operators $O_{6,8}$ to the hadronic
 matrix elements vanish, i.e., the terms that are related to $a_{6,8}$
 disappear from the decay amplitudes for case II. As to the parameters
 $H(BM_{1},M_{2})$ and $H^{\prime}(BM_{1},M_{2})$ in $a_{i,II}$, they
 are defined as
 \begin{mathletters}
 \begin{eqnarray}
  H(BP,V)&=&\frac{f_{B}f_{P}}{m_{B}^{2}F_{1}^{B{\to}P}}
  {\int}_{0}^{1} d{\xi} {\int}_{0}^{1} dx {\int}_{0}^{1} dy
   \frac{{\Phi}_{B}({\xi})}{\xi}
   \frac{{\Phi}_{V}(x)}{\bar{x}} %\nonumber \\ & &
   \Big[ \frac{{\Phi}_{P}(y)}{\bar{y}}
       + \frac{2 {\mu}_{P}}{m} \frac{\bar{x}}{x}
         \frac{{\Phi}_{P}^{p}(y)}{\bar{y}} \Big],
   \\
  H^{\prime}(BP,V)&=& - \frac{f_{B}f_{P}}{m_{B}^{2}F_{1}^{B{\to}P}}
  {\int}_{0}^{1} d{\xi} {\int}_{0}^{1} dx {\int}_{0}^{1} dy
   \frac{{\Phi}_{B}({\xi})}{\xi}
   \frac{{\Phi}_{V}(x)}{x} %\nonumber \\ & &
   \Big[ \frac{{\Phi}_{P}(y)}{\bar{y}}
       + \frac{2 {\mu}_{P}}{m} \frac{x}{\bar{x}}
         \frac{{\Phi}_{P}^{p}(y)}{\bar{y}} \Big].
 \end{eqnarray}
 \label{eq:hard-spectator-v}
 \end{mathletters}

 Now we would like to make some comments on $a_{i}$
 \begin{enumerate}
 \item It was shown in \cite{dyz} that at leading order approximation
       $a_{i}$ are renormalization scale-independent, i.e.,
       \begin{equation}
       \frac{d \ a_{i,I}}{d \ {\ln}{\mu}}=0 \ \ (i{\ne}6,8)
       \ \ \ \ \ \ \text{and} \ \ \ \ \ \
       \frac{d \ a_{i,I} r_{\chi}}{d \ {\ln}{\mu}}=0 \ \ (i=6,8)
       \label{eq:scale}
       \end{equation}
       where $r_{\chi}$ is the chirally enhanced factor, e.g.,
       $r_{\chi}^{K}(\mu)=\frac{2m_{K^{+}}^{2}}{m_{b}(\mu)
       [m_{u}(\mu)+m_{s}(\mu)]}$ for the decay $B^{0}{\to}K^{+}{\rho}^{-}$.
       It is formally ${\cal O}({\Lambda}_{QCD}/m_{b})$ power suppressed,
       but numerically large. The deviation of the data in Table 
       \ref{tab2} from Eq.(\ref{eq:scale}) can be reduced further when the
       higher order radiative corrections to the hadronic matrix elements
       are included. So in the following calculations, we will fix
       ${\mu}=m_{b}$. In addition, it has been proved that $a_{i}$ are
       free of gauge dependence. These two points guarantee that the decay
       amplitudes are physical.
 \item From Eq.(\ref{eq:ai}) we can see that, in the heavy quark limit,
       when the corrections at the order of ${\alpha}_{s}$ and $\alpha$
       are neglected, we return to the NF approximation.
 \item ``Nonfactorizable'' effects appear at the order of ${\alpha}_{s}$
       or/and ${\Lambda}_{QCD}/m_{b}$. This reflects the fact that the
       final state interactions are either dominated by hard processes or
       power suppressed. Therefore the strong phases for the class-I
       ($a_{1}$ dominant) decays are small and furthermore perturbatively
       calculable in the heavy quark limit. Moreover, from the data in
       Table \ref{tab2}, we find that the ``nonfactorizable'' effects
       contribute a large imaginary part to $a_{2,I}$, i.e., the class-II
       ($a_{2}$ dominant) decays, especially the CP asymmetries, might be
       sensitive to the ``nonfactorizable'' corrections.
 \end{enumerate}
 
 \subsection{The annihilation parameters $b_{i}$}
 \label{sec24}
 The parameters of $b_{i}$ in Eq.(\ref{eq:decay-a}) correspond to weak
 annihilation contributions. Now we give their expressions, which are
 analogous to those in \cite{0104110}:
 \begin{mathletters}
 \begin{eqnarray}
  & &b_{1}(M_{1},M_{2}) = \frac{C_{F}}{N_{c}^{2}} C_{1}
                           A_{1}^{i}(M_{1},M_{2}),
     \ \ \ \ \ \ \ \ \
     b_{2}(M_{1},M_{2}) = \frac{C_{F}}{N_{c}^{2}} C_{2}
                           A_{1}^{i}(M_{1},M_{2}), \\
  & &b_{3}(M_{1},M_{2}) = \frac{C_{F}}{N_{c}^{2}} \Big\{ C_{3}
                           A_{1}^{i}(M_{1},M_{2})
                   + C_{5} A_{3}^{i}(M_{1},M_{2})
             + \Big[ C_{5} + N_{c} C_{6} \Big]
                           A_{3}^{f}(M_{1},M_{2}) \Big\}, \\
  & &b_{4}(M_{1},M_{2}) = \frac{C_{F}}{N_{c}^{2}} \Big\{ C_{4}
                           A_{1}^{i}(M_{1},M_{2})
                   + C_{6} A_{2}^{i}(M_{1},M_{2}) \Big\}, \\
  & &b_{3}^{ew}(M_{1},M_{2}) = \frac{C_{F}}{N_{c}^{2}} \Big\{ C_{9}
                           A_{1}^{i}(M_{1},M_{2})
                   + C_{7} A_{3}^{i}(M_{1},M_{2})
             + \Big[ C_{7} + N_{c} C_{8} \Big]
                           A_{3}^{f}(M_{1},M_{2}) \Big\}, \\
  & &b_{4}^{ew}(M_{1},M_{2}) = \frac{C_{F}}{N_{c}^{2}} \Big\{ C_{10}
                           A_{1}^{i}(M_{1},M_{2})
                   + C_{8} A_{2}^{i}(M_{1},M_{2}) \Big\}.
 \end{eqnarray}
 \label{eq:bi}
 \end{mathletters}
 Here the current-current annihilation parameters $b_{1,2}(M_{1},M_{2})$
 arise from the hadronic matrix elements of the effective operators
 $O_{1,2}$, the QCD penguin annihilation parameters $b_{3,4}(M_{1},M_{2})$
 from $O_{3-6}$, and the electroweak penguin annihilation parameters
 $b_{3,4}^{ew}(M_{1},M_{2})$ from $O_{7-10}$. The parameters of $b_{i}$
 are closely related to the final states; they can also be divided into
 two different cases according to the final states. Case I is that $M_{1}$
 is a vector meson, and $M_{2}$ is a pseudoscalar meson (here $M_{1}$ and
 $M_{2}$ are tagged in Fig.2). Case II is that $M_{1}$ corresponds to a
 pseudoscalar meson, and $M_{2}$ corresponds to a vector meson. For
 case I, the definitions of $A^{i,f}_{k}(M_{1},M_{2})$
 in Eq.(\ref{eq:bi}) are
 \begin{mathletters}
 \begin{eqnarray}
 & &A^{f}_{1}(V,P) = 0, \ \ \ \ \ \ \ \ \ \ \ \
    A^{f}_{2}(V,P) = 0,
 \label{eq:bi-fvp-1} \\
 & &A^{f}_{3}(V,P) = {\pi}{\alpha}_{s} {\int}_{0}^{1} dx {\int}_{0}^{1} dy
            {\Phi}_{V}(x) {\Phi}_{P}^{p}(y) \frac{2 {\mu}_{P}}{m}
             \frac{2 (1+\bar{x})}{{\bar{x}}^{2}y},
 \label{eq:bi-fvp-2} \\
 & &A^{i}_{1}(V,P) = {\pi}{\alpha}_{s}
            {\int}_{0}^{1} dx {\int}_{0}^{1} dy
            {\Phi}_{V}(x) {\Phi}_{P}(y)
       \Big[ \frac{1}{y(1-x\bar{y})}
           + \frac{1}{{\bar{x}}^{2}y} \Big],
 \label{eq:bi-ivp-1} \\
 & &A^{i}_{2}(V,P) = - {\pi}{\alpha}_{s}
            {\int}_{0}^{1} dx {\int}_{0}^{1} dy
            {\Phi}_{V}(x) {\Phi}_{P}(y)
       \Big[ \frac{1}{\bar{x}(1-x\bar{y})}
           + \frac{1}{\bar{x}y^{2}} \Big],
 \label{eq:bi-ivp-2} \\
 & &A^{i}_{3}(V,P) = {\pi}{\alpha}_{s}
            {\int}_{0}^{1} dx {\int}_{0}^{1} dy
            {\Phi}_{V}(x) {\Phi}_{P}^{p}(y)
             \frac{2 {\mu}_{p}}{m}
             \frac{2\bar{y}}{\bar{x}y(1-x\bar{y})}.
 \label{eq:bi-ivp-3}
 \end{eqnarray}
 \label{eq:bi-ivp}
 \end{mathletters}
 \begin{mathletters}
 For case-II,
 \begin{eqnarray}
 & &A^{f}_{1}(P,V)=0, \ \ \ \ \ \ \ \ \ \ \ \
    A^{f}_{2}(P,V)=0,
 \label{eq:bi-fpv-1} \\
 & &A^{f}_{3}(P,V)= - {\pi}{\alpha}_{s} {\int}_{0}^{1} dx
            {\int}_{0}^{1} dy {\Phi}_{P}^{p}(x) {\Phi}_{V}(y)
             \frac{2 {\mu}_{P}}{m} \frac{2 (1+y)}{\bar{x}y^{2}},
 \label{eq:bi-fpv-2} \\
 & &A^{i}_{1}(P,V) = {\pi}{\alpha}_{s}
            {\int}_{0}^{1} dx {\int}_{0}^{1} dy
            {\Phi}_{P}(x) {\Phi}_{V}(y)
       \Big[ \frac{1}{y(1-x\bar{y})}
           + \frac{1}{{\bar{x}}^{2}y} \Big],
 \label{eq:bi-ipv-1} \\
 & &A^{i}_{2}(P,V) = - {\pi}{\alpha}_{s}
            {\int}_{0}^{1} dx {\int}_{0}^{1} dy
            {\Phi}_{P}(x) {\Phi}_{V}(y)
       \Big[ \frac{1}{\bar{x}(1-x\bar{y})}
           + \frac{1}{\bar{x}y^{2}} \Big],
 \label{eq:bi-ipv-2}  \\
 & &A^{i}_{3}(P,V) = {\pi}{\alpha}_{s}
            {\int}_{0}^{1} dx {\int}_{0}^{1} dy
            {\Phi}_{P}^{p}(x) {\Phi}_{V}(y)
             \frac{2 {\mu}_{p}}{m}
             \frac{2x}{\bar{x}y(1-x\bar{y})}.
 \label{eq:bi-ipv-3}
 \end{eqnarray}
 \label{eq:bi-ipv}
 \end{mathletters}
 Here our notation and convention are the same as those in \cite{0104110}.
 The superscripts $i$ and $f$ on $A^{i,f}$ correspond to the contributions
 from Fig.2(a-b) and Fig.2(c-d), respectively. The subscripts $k=1,2,3$ on
 $A^{i,f}_{k}$ refer to the Dirac structures $(V-A){\otimes}(V-A)$,
 $(V-A){\otimes}(V+A)$ and $(-2)(S-P){\otimes}(S+P)$, respectively.
 ${\Phi}_{V}(x)$ denotes the leading-twist LCDAs of a vector meson, and
 ${\Phi}_{P}(x)$ and ${\Phi}_{P}^{p}(x)$ denote twist-2 and twist-3 LCDAs
 of a pseudoscalar meson, respectively\footnote{
 The expression for $A_{3}^{i}$ is different from that in \cite{0012152}.
 This difference originates from the two different ways of dealing with
 the annihilation contribution from the twist-3 LCDAs. Cheng and Yang 
 calculated them in coordinate space, and we take the projection in 
 momentum space given in \cite{0104110}. We made some investigation of 
 this problem in \cite{dyz}, and discussed it with Beneke et al. The point
 lies in how to deal with the surface terms in the integral. In addition,
 we found that, when the annihilation contributions from the twist-3 LCDAs
 are considered for $B{\to}PP$ decays in coordinate space, not only 
 logarithmic divergence but also linear and quadratic divergences appear,
 which is a very serious problem. So we use the method given in
 \cite{0104110}, which gives only a logarithmic divergence even for
 $B{\to}PP$ decays.}.

 Note that assuming $SU(3)$ flavor symmetry and symmetric (under
 $x{\leftrightarrow}\bar{x}$) LCDAs of light mesons, we have
 $A_{1}^{i}=-A_{2}^{i}$. In this approximation the weak annihilation
 contributions (for case I) can be parametrized as
 \begin{mathletters}
 \begin{eqnarray}
 A_{1}^{i}(V,P) &{\simeq}& 18{\pi}{\alpha}_{s}\Big(X_{A}-4
             +\frac{{\pi}^{2}}{3}\Big), \\
 A_{3}^{i}(V,P) &{\simeq}& {\pi}{\alpha}_{s}r_{\chi}\bigg[
              2{\pi}^{2}-6\Big(X_{A}^{2}+2X_{A}\Big)\bigg], \\
 A_{3}^{f}(V,P) &{\simeq}& 6{\pi}{\alpha}_{s}r_{\chi}
             \Big(2X_{A}^{2}-X_{A}\Big),
 \end{eqnarray}
 \label{eq:aif-kvp}
 \end{mathletters}
 where $X_{A}={\int}^{1}_{0}dx/x$ parametrizes the divergent endpoint
 integrals. We can get similar forms to Eq.(\ref{eq:aif-kvp}) for case II,
 but with $A_{3}^{f}(P,V)=-A_{3}^{f}(V,P)$. In our calculation, we will
 treat $X_{A}$ as a phenomenological parameter, and take the same value
 for all annihilation terms, although this approximation is crude and
 there is no known physical argument for justifying this assumption. We
 shall see below that $X_{A}$ gives large uncertainties in the theoretical
 prediction.
 
 \section{Input parameters}
 \label{sec3}
 The QCDF expressions for the hadronic matrix elements are written as the
 product of hard-scattering kernels and LCDAs of mesons. The hard
 scattering kernels are perturbatively calculable, while the soft and
 nonperturbative effects are incorporated into two kinds of universal
 parameter: the form factors and the LCDAs of mesons. The decay amplitudes
 are also related to the CKM matrix elements, the masses of quarks and
 mesons, various decay constants of mesons, and so on. We will specify
 them in the following discussions.

 \subsection{The CKM matrix elements}
 \label{sec31}
 The CKM matrix is described by four independent parameters. Using the
 Wolfenstein parametrization \cite{Wolfenstein}, it can be expressed as
 \begin{equation}
 V_{CKM}=\left( \begin{array}{ccc}
     1-{\lambda}^{2}/2
  &    {\lambda}
  &   A{\lambda}^{3}({\rho}-i{\eta})  \\
      -{\lambda}
  &  1-{\lambda}^{2}/2
  &   A{\lambda}^{2}                  \\
      A{\lambda}^{3}(1-{\rho}-i{\eta})
  &  -A{\lambda}^{2}
  &  1  \end{array} \right) + {\cal O}({\lambda}^{4}).
 \label{eq:ckm}
 \end{equation}
 Two of them are well-determined: $A=0.819{\pm}0.040$ and
 ${\lambda}=0.2237{\pm}0.0033$ \cite{0012308}. Recently M. Ciuchini
 revised the values of ${\rho}$ and ${\eta}$ in \cite{0112133}
 $\bar{\rho}=0.218{\pm}0.038$, $\bar{\eta}=0.316{\pm}0.040$, and
 ${\gamma}=(55.5{\pm}6.2)^{\circ}$. This gives ${\rho}=0.224{\pm}0.039$
 and ${\eta}=0.324{\pm}0.039$. A. H\"{o}cker, et al. \cite{0104062}
 advocate another approach to a global CKM matrix analysis, using the
 frequentist statistics named {\it R}fit; their best fit results are
 $A=0.83{\pm}0.07$, ${\lambda}=0.222{\pm}0.004$,
 $\bar{\rho}=0.21{\pm}0.12$, $\bar{\eta}=0.38{\pm}0.11$ and
 ${\gamma}=(62{\pm}15)^{\circ}$ at $95\%$ confidence level. H. Lacker and
 M. Neubert et al. suggest that it is possible to derive constraints on
 ${\gamma}$ from a global {\it R}fit analysis of various CP-averaged
 branching ratios $B{\to}{\pi}{\pi},{\pi}K$; their best fit of the QCDF
 theory to the data yields $(\bar{\rho},\bar{\eta})=(0.05,0.381)$ with
 ${\chi}^{2}/n_{dof}=0.46$ \cite{0110301}, and correspondingly
 ${\gamma}{\simeq}83^{\circ}$. In our calculation, we will take the
 central values of the results in \cite{0112133} as input parameters.
 At the same time, we will also give estimations based on the results 
 in \cite{0110301}.
 
 \subsection{Quark masses}
 \label{sec32}
 The masses of quarks appear in our calculations in two different ways.
 The pole masses of quarks arise from the loop integration over the
 virtual internal quarks in penguin correction diagrams. They contribute
 to the penguin parameters $P_{M,i}^{p}$ and $P_{M,i}^{p,ew}$ in
 Eq.(\ref{eq:penguin}) in terms of $G_{M}(s_{q})$ and 
 ${\hat{G}}_{M}(s_{q})$ \cite{0104110}, where
 $s_{q}=m_{q}^{2}/m_{b}^{2}$. We fix them as
 \begin{equation}
  m_{u}=m_{d}=m_{s}=0,   \ \ \ \
  m_{c}=1.45 \text{GeV}, \ \ \ \
  m_{b}=4.6 \text{GeV}.
 \label{eq:mass-1}
 \end{equation}
 The other type of quark mass is the running quark mass which is
 renormalization scale-dependent. It appears in terms of the chirally
 enhanced factor $r_{\chi}$ which arises from the hadronic matrix elements
 of $(S+P){\otimes}(S-P)$ operators through the equations of motion, and
 the twist-3 LCDAs of the pseudoscalar mesons. Estimations of the running
 quark masses within the QCD sum rules approach are collected in 
 \cite{0010175}. Here we would like to  use the central values of Particle
 Data Group 2000 data \cite{PDG2000} for discussion.
 \begin{mathletters}
 \begin{eqnarray}
 & &{\overline{m}}_{b}({\overline{m}}_{b})=4.2\text{GeV},
    \ \ \ \ \ \ \ \ \ \ \
    {\overline{m}}_{s}(2\text{GeV})=122.5 \text{MeV},\\
 & &{\overline{m}}_{d}(2\text{GeV})=6 \text{MeV},
    \ \ \ \ \ \ \ \ \
    {\overline{m}}_{u}(2\hbox{GeV})=3 \text{MeV}.
 \end{eqnarray}
 \label{eq:mass-2}
 \end{mathletters}
 Using the renormalization group equation, we can get their corresponding
 values at the scale ${\mu}={\cal O}(m_{b})$. In addition, we would like
 to point out that, because the running masses of light quarks have large
 uncertainties, we will take the $r_{\chi}^{{\eta}^{({\prime})}}
 (1-f^{u}_{{\eta}^{({\prime})}}/f^{s}_{{\eta}^{({\prime})}})
 {\simeq}r_{\chi}^{\pi}{\simeq}r_{\chi}^{K}{\equiv}r_{\chi}$
 approximation for simplicity in our numerical calculations.

 \subsection{The form factors and decay constants}
 \label{sec33}
 Now let us parametrize the hadronic matrix elements of
 ${\langle}PV{\vert}O_{i}{\vert}B{\rangle}_{f}$ in Eq.(\ref{eq:decay-f}).
 After Fierz reordering, with the NF assumption \cite{bsw}, they can be
 written as
 \begin{equation}
 {\langle}PV{\vert}O_{i}{\vert}B{\rangle} =
     Z_{1} {\langle}P{\vert}J^{\mu}{\vert}0{\rangle}
           {\langle}V{\vert}J_{\mu}{\vert}B{\rangle}
   + Z_{2} {\langle}V{\vert}J^{{\prime}{\mu}}{\vert}0{\rangle}
           {\langle}P{\vert}J^{\prime}_{\mu}{\vert}B{\rangle},
 \label{eq:factorization}
 \end{equation}
 where $J^{\mu}$ and $J^{{\prime}{\mu}}$ are hadronic currents and
 $Z_{1,2}$ are the corresponding coefficients. The hadronic current
 matrix elements are defined as follows \cite{bsw}:
 \begin{mathletters}
 \begin{eqnarray}
 {\langle}P(l){\vert}\bar{q}{\gamma}^{\mu}
        {\gamma}_{5}q{\vert}0{\rangle}
    &=& - i f_{P} l^{\mu}, \ \ \ \ \ \ \ \ \ \ \ \ \
 {\langle}V(l,{\epsilon}){\vert}\bar{q}{\gamma}^{\mu}
        q{\vert}0{\rangle}
     = f_{V} m_{V} {\epsilon}^{{\ast}{\nu}}, \\
 {\langle}P(l){\vert}\bar{q}{\gamma}^{\mu}
       (1-{\gamma}_{5})q{\vert}B{\rangle}
    &=& \Big[ p_{B}^{\mu} + l^{\mu}
     - \frac{m_{B}^{2}-m_{P}^{2}}{k^{2}} k^{\mu}
       \Big] F_{1}(k^{2})
     + \frac{m_{B}^{2}-m_{P}^{2}}{k^{2}} k^{\mu} F_{0}(k^{2}), \\
 {\langle}V(l,{\epsilon}){\vert}\bar{q}{\gamma}_{\mu}
       (1-{\gamma}_{5})q{\vert}B{\rangle}
    &=& {\epsilon}_{{\mu}{\nu}{\alpha}{\beta}}{\epsilon}^{{\ast}{\nu}}
       p_{B}^{\alpha} l^{\beta} \frac{2V(k^{2})}{m_{B}+m_{V}}
     + i \frac{2m_{V}({\epsilon}^{\ast}{\cdot}k)}{k^{2}}
       k_{\mu} A_{0}(k^{2}) \nonumber \\
    &+& i {\epsilon}^{\ast}_{\mu} ( m_{b}+m_{V} ) A_{1}(k^{2})
     - i \frac{{\epsilon}^{\ast}{\cdot}k}{m_{b}+m_{V}}
          ( p_{B} + l )_{\mu} A_{2}(k^{2}) \nonumber \\
    &-& i \frac{2m_{V}({\epsilon}^{\ast}{\cdot}k)}{k^{2}}
       k_{\mu} A_{3}(k^{2}),
 \end{eqnarray}
 \label{eq:hadronic-j}
 \end{mathletters}
 where $k=p_{B}-l$, and ${\epsilon}^{\ast}$ denotes the polarization
 vector of the vector meson $V$. $f_{P}$ and $f_{V}$ are decay
 constants, and $F_{0,1}(k^{2})$, $V(k^{2})$, and $A_{0,1,2,3}(k^{2})$ 
 are form factors. In addition, at the poles $k^{2}=0$, we have
 \begin{mathletters}
 \begin{eqnarray}
  & &F_{0}(0)=F_{1}(0), \ \ \ \ \ \ \ \ \ \ \ \ \ \
     A_{0}(0)=A_{3}(0), \\
  & &2 m_{V} A_{3}(0) = (m_{B}+m_{V}) A_{1}(0) - (m_{B}-m_{V}) A_{2}(0).
 \label{eq:conditions}
 \end{eqnarray}
 \end{mathletters}
 In the QCDF framework, they are all nonperturbative quantities, and
 appear as universal input parameters. As a good approximation, we take
 these form factors at $k^{2}=0$ in our calculations.

 Because the flavor octet mixes with the flavor singlet in the $SU(3)$
 quark representation of mesons, the corresponding hadronic parameters
 are difficult to determine. Here we assume ideal mixing between $\omega$
 and $\phi$, i.e., ${\omega}=(u\bar{u}+d\bar{d})/{\sqrt{2}}$ and
 ${\phi}=s\bar{s}$. As to $\eta$ and ${\eta}^{\prime}$, we follow the
 convention in \cite{ali,9709408,9710268}: the two-mixing-angle formula is
 applied for the decay constants and form factors, and the charm quark
 content in $\eta$ and ${\eta}^{\prime}$ is assumed to be negligible:
 \begin{mathletters}
 \begin{eqnarray}
 & &{\langle}0{\vert}\bar{q}{\gamma}_{\mu}{\gamma}_{5}q
           {\vert}{\eta}^{(\prime)}(p){\rangle}
          =if^{q}_{{\eta}^{(\prime)}}p_{\mu},
    \ \ \ \ \ \ \ (q=u,d,s) \\
 & &\frac{{\langle}0{\vert}\bar{u}{\gamma}_{5}u
                    {\vert}{\eta}^{(\prime)}{\rangle}}
         {{\langle}0{\vert}\bar{s}{\gamma}_{5}s
                    {\vert}{\eta}^{(\prime)}{\rangle}}
  = \frac{f^{u}_{{\eta}^{(\prime)}}}{f^{s}_{{\eta}^{(\prime)}}},
    \ \ \ \ \ \
          {\langle}0{\vert}\bar{s}{\gamma}_{5}s
                    {\vert}{\eta}^{(\prime)}{\rangle}
  = -i\frac{m_{{\eta}^{(\prime)}}^{2}}{2m_{s}}
   (f^{s}_{{\eta}^{(\prime)}}-f^{u}_{{\eta}^{(\prime)}}),\\
 & &f^{u}_{\eta}=  \frac{f_{8}}{\sqrt{6}}{\cos}{\theta}_{8}
                -  \frac{f_{0}}{\sqrt{3}}{\sin}{\theta}_{0},
    \ \ \ \ \ \ \ \
    f^{s}_{\eta}=-2\frac{f_{8}}{\sqrt{6}}{\cos}{\theta}_{8}
                -  \frac{f_{0}}{\sqrt{3}}{\sin}{\theta}_{0},\\
 & &f^{u}_{{\eta}^{\prime}}=  \frac{f_{8}}{\sqrt{6}}{\sin}{\theta}_{8}
                           +  \frac{f_{0}}{\sqrt{3}}{\cos}{\theta}_{0},
    \ \ \ \ \ \ \ \
    f^{s}_{{\eta}^{\prime}}=-2\frac{f_{8}}{\sqrt{6}}{\sin}{\theta}_{8}
                           +  \frac{f_{0}}{\sqrt{3}}{\cos}{\theta}_{0},\\
 & &F_{0,1}^{B{\eta}}=F_{0,1}^{B{\pi}}\Big(
         \frac{{\cos}{\theta}_{8}}{\sqrt{6}}
        -\frac{{\sin}{\theta}_{0}}{\sqrt{3}}\Big), \ \ \ \ \
    F_{0,1}^{B{\eta}^{\prime}}=F_{0,1}^{B{\pi}}\Big(
         \frac{{\sin}{\theta}_{8}}{\sqrt{6}}
        +\frac{{\cos}{\theta}_{0}}{\sqrt{3}}\Big).
 \end{eqnarray}
 \label{eq:eta-etap}
 \end{mathletters}
 We take the mixing angles as ${\theta}_{8}=-22.2^{\circ}$ and
 ${\theta}_{0}=-9.1^{\circ}$. For the $B{\to}K$ transition form factors
 we use the $SU(3)$ flavor symmetry approximation
 \begin{equation}
 \frac{f_{\pi}}{f_{K}}{\approx}\frac{F_{0,1}^{B{\pi}}(0)}{F_{0,1}^{BK}(0)}.
 \end{equation}

 The decay constants and form factors are nonperturbative parameters; they
 are available from the experimental data and/or estimated with 
 well-founded theories, such as lattice calculations, QCD sum rules etc.
 Now we take their values in our calculations as \cite{ali,PDG2000,%
 9506398,0001297,9802394,9401277}:
 \begin{eqnarray*}
 & &f_{\pi}=131\text{MeV},    \ \ \ \ \ \
    f_{K}=160\text{MeV},      \ \ \ \ \ \
    f_{K^{\ast}}=214\text{MeV}, \\
 & &f_{\rho}=210\text{MeV},   \ \ \ \ \ \
    f_{\omega}=195\text{MeV}, \ \ \ \ \ \
    f_{\phi}=233\text{MeV},     \\
 & &f_{0}=157\text{MeV}, \ \ \ \ \ \
    f_{8}=168\text{MeV}, \ \ \ \ \ \
    f_{B}=180\text{MeV},   \\
 & &F^{B{\pi}}_{0,1}(0)=0.28{\pm}0.05, \ \ \ \ \ \ \ \ \ \ \ \
    A^{BK^{\ast}}_{0}(0)=0.39{\pm}0.10,\\
 & &A^{B{\rho}}_{0}(0)=0.30{\pm}0.05,  \ \ \ \ \ \ \ \ \ \ \ \
    A^{B{\omega}}_{0}(0)=0.30{\pm}0.05.
 \end{eqnarray*}

 \subsection{The LCDAs of mesons}
 \label{sec34}
 The LCDAs of mesons are also basic input parameters in the QCDF
 formula Eq.(\ref{eq:qcdf}). The LCDAs of light pseudoscalar mesons
 are defined as
 \cite{LCDAs}:
 \begin{eqnarray}
 & &{\langle}P(k){\vert}\bar{q}(z_{2})q(z_{1}){\vert}0{\rangle}
    \nonumber \\
 &=&\frac{if_{P}}{4}{\int}_{0}^{1}dx \
    e^{i(xk{\cdot}z_{2}+\bar{x}k{\cdot}z_{1})} \Big\{
    k\!\!\!\slash {\gamma}_{5} {\Phi}_{P}(x)
    - {\mu}_{P} {\gamma}_{5} \Big[ {\Phi}_{P}^{p}(x)
    - {\sigma}_{{\mu}{\nu}}k^{\mu}z^{\nu}
       \frac{{\Phi}_{P}^{\sigma}(x)}{6} \Big] \Big\},
 \label{eq:LCDAs-1}
 \end{eqnarray}
 where $f_{P}$ is a decay constant. $z=z_{2}-z_{1}$, $\bar{x}=1-x$, and
 ${\mu}_{P}=\frac{m_{P}^{2}}{[m_{1}(\mu)+m_{2}(\mu)]}$ (here $m_{1}$
 and $m_{2}$ are running masses of the valence quarks of the
 pseudoscalar meson ($P$) with mass $m_{P}$.), ${\Phi}_{P}(x)$ is the
 leading twist LCDAs, and ${\Phi}_{P}^{p}(x)$, ${\Phi}_{P}^{\sigma}(x)$
 are twist-3 LCDAs of the meson. We shall use the asymptotic forms of
 the LCDAs for the following discussion:
 \begin{equation}
  {\Phi}_{P}(x)=6x\bar{x}, \ \ \ \ \ \
  {\Phi}_{P}^{p}(x)=1, \ \ \ \ \ \
  {\Phi}_{P}^{\sigma}(x)=6x\bar{x}.
 \label{eq:LCDAs-2}
 \end{equation}

 The twist-2 LCDAs of the vector mesons are defined as
 \cite{9401277,9810475}:
 \begin{mathletters}
 \begin{eqnarray}
 {\langle}0{\vert}\bar{q}(0){\sigma}_{{\mu}{\nu}}q(z)
           {\vert}V(k,{\lambda}){\rangle}&=&
  i ( {\epsilon}_{\mu}^{\lambda}k_{\nu}
    - {\epsilon}_{\nu}^{\lambda}k_{\mu} )
      f_{V}^{\bot} {\int}_{0}^{1} dx \
      e^{-ixk{\cdot}z} {\Phi}_{V}^{\bot}(x), \\
 {\langle}0{\vert}\bar{q}(0){\gamma}_{\mu}q(z)
           {\vert}V(k,{\lambda}){\rangle}&=&
    k_{\mu} \frac{{\epsilon}^{\lambda}{\cdot}z}{k{\cdot}z}
    f_{V} m_{V} {\int}_{0}^{1} dx \
    e^{-ixk{\cdot}z} {\Phi}_{V}^{\|}(x),
 \end{eqnarray}
 \label{eq:LCDAs-3}
 \end{mathletters}
 where ${\epsilon}$ is a polarization vector, and for longitudinally
 polarized mesons ${\epsilon}_{\|}=k/m_{V}$. ${\Phi}_{V}^{\|}(x)$ and
 ${\Phi}_{V}^{\bot}(x)$ describe the quark distributions of transversely
 and longitudinally polarized mesons. In our calculations, the
 contributions from ${\Phi}_{V}^{\bot}(x)$ are power suppressed and
 hence can be neglected, i.e., we take the following approximation:
 \begin{equation}
 {\Phi}_{V}(x)={\Phi}_{V}^{\|}(x)=6x\bar{x}.
 \label{eq:LCDAs-4}
 \end{equation}

 For the wave function of the B meson, we take the form in
 \cite{0004173,0004213,bauer89}:
 \begin{equation}
 {\Phi}_{B}({\xi})=N_{B}{\xi}^{2}(1-{\xi})^{2}{\exp}\Big[
  -\frac{m_{B}^{2}{\xi}^{2}}{2{\omega}_{B}^{2}}\Big],
 \label{eq:B-function}
 \end{equation}
 where $N_{B}$ is the normalization constant. ${\Phi}_{B}({\xi})$ is
 peaked around ${\xi}{\approx}0.1$ with ${\omega}_{B}=0.4\text{GeV}$.

 With these LCDAs of the mesons, we still cannot get the parameters
 $a_{i}$ and $b_{i}$ at once, because we will encounter endpoint
 divergence in dealing with the integrals Eq.(\ref{eq:hard-spectator-v}), 
 Eq.(\ref{eq:bi-ivp}), and Eq.(\ref{eq:bi-ipv}). If the transverse
 momentum $k_{\bot}$ of the partons and the Sudakov resummation can be
 taken into account consistently in the QCDF approach, the above-mentioned
 integrals might be convergent. But now we have to treat these divergent
 integrals as phenomenological parameters:
 \begin{equation}
  X = {\int}_{0}^{1}\frac{dx}{x}
    = {\ln}\frac{m_{B}}{\overline{\Lambda}}
    + {\varrho} e^{-i{\phi}},
 \label{eq:divergence}
 \end{equation}
 where $\varrho$ can vary from 0 to 6, ${\phi}$ is an arbitrary phase,
 and $0^{\circ}{\le}{\phi}{\le}360^{\circ}$. These notations are almost
 the same as those in \cite{0104110,0012152}. In our calculation, $X_{H}$
 and $X_{A}$ denote the endpoint divergent integrals from hard spectator
 scattering Eq.(\ref{eq:hard-spectator-v}), and weak annihilations
 Eq.(\ref{eq:bi-ivp}), Eq.(\ref{eq:bi-ipv}), respectively. We take
 $\overline{\Lambda}={\Lambda}_{QCD}$ and
 ${\varrho}e^{-i{\phi}}=i{\pi}$ as default values.

 \section{Branching ratios}
 \label{sec4}
 The branching ratios of charmless decays $B{\to}PV$ in the B meson
 rest frame can be written as
 \begin{equation}
 {\cal B}r(B{\to}PV)= \frac{{\tau}_{B}}{8{\pi}}
  \frac{{\vert}p{\vert}}{m_{B}^{2}}{\vert}{\cal A}(B{\to}PV){\vert}^{2}.
 \label{eq:br-1}
 \end{equation}
 In our calculation, we take ${\tau}_{B^{0}}=1.548\text{ps}$,
 ${\tau}_{B^{\pm}}=1.653\text{ps}$, and
 \begin{equation}
 {\vert}p{\vert}=\frac{\sqrt{[m_{B}^{2}-(m_{P}+m_{V})^{2}]
       [m_{B}^{2}-(m_{P}-m_{V})^{2}]}}{2m_{B}}.
 \label{eq:br-2}
 \end{equation}
 Since QCD factorization works in the heavy quark limit, the above masses
 of light mesons should be taken as zero for consistency.

 The experimental measurements are collected in Table \ref{tab3}. Our
 numerical results for CP-averaged branching ratios for $B{\to}PV$
 are listed in Table \ref{tab4} and Table \ref{tab5}, calculated at
 the scale ${\mu}=m_{b}$ for two choices of CKM matrix
 elements: (1) $A=0.819$, ${\lambda}=0.2237$, $\bar{\rho}=0.218$,
 and $\bar{\eta}=0.316$ \cite{0112133}; (2) $A=0.83$, ${\lambda}=0.222$,
 $\bar{\rho}=0.05$, and $\bar{\eta}=0.381$ \cite{0110301}. The other
 related parameters are taken at their default values. In the tables,
 ${\cal B}r$ stands for the values that are calculated without including
 weak annihilation contributions within the NF framework;
 ${{\cal B}r}^{f}{\propto}{\vert}{\cal A}^{f}{\vert}^{2}$ and
 ${{\cal B}r}^{f+a}{\propto}{\vert}{\cal A}^{f}+{\cal A}^{a}{\vert}^{2}$
 are estimated with the QCDF approach where ${\cal A}^{a}$ denotes
 the annihilation contributions. Since the above two choices of CKM matrix
 elements are nearly equivalent to two choices of the angle $\gamma$, many
 of the CP-averaged branching ratios are similar to each other for these
 two sets of CKM matrix elements, as can be seen from Table \ref{tab4} and
 Table \ref{tab5}. Only the decay channels that have large interference
 between tree and penguin contributions are sensitive to the angle
 $\gamma$, such as ${\overline{B}}^{0}{\to}{\pi}^{+}K^{{\ast}-}$ and
 $B^{-}{\to}{\pi}^{0}K^{{\ast}-}$. It is thus possible to extract $\gamma$
 from these decay channels. However, we must consider the uncertainties
 due to the variations of various parameters, such as form factors, and
 the model dependence of the parametrization of chirally enhanced
 hard-spectator contributions and annihilation contributions, and so on.
 Nevertheless, it is still possible to obtain some information on $\gamma$
 with detailed analysis. The readers may notice that some numerical
 results in Table \ref{tab4} and Table \ref{tab5} are inconsistent with
 the experimental measurements listed in Table \ref{tab3}. The
 inconsistency is not serious because we use default values of the
 parameters for the tables and do not consider the uncertainties from
 the input parameters. In fact, we will see that for appropriate regions
 of parameters predictions from the QCDF approach are in agreement with
 current measurements for most of the $B{\to}PV$ decays.

 From the experience in $B \rightarrow PP$ analysis \cite{0108141}, we
 know that the main theoretical uncertainties of branching ratios come
 from CKM matrix elements, form factors, and weak annihilation
 contributions. It is easy to imagine the importance of CKM matrix
 elements and form factors, but for weak annihilation topologies the
 importance was first noticed recently in \cite{0004173} within the  
 perturbative QCD approach. However, within the QCDF framework, the
 annihilation topologies introduce endpoint divergence, which violates
 factorization; in Ref. \cite{0104110}, the authors phenomenologically
 parametrize the endpoint divergence integral as $X_A$. In this work, we
 will follow their way of estimating the annihilation contributions,
 although it will introduce model dependence and numerical uncertainties.

 We know that the leading power of the annihilation contribution is always
 $X_A^2$ for $B{\to}PP$ decay amplitudes. But for $B{\to}PV$, the case is
 different: According to the formulas in Sec. \ref{sec24}, $b_3^{ew}$ and
 $b_4^{ew}$ are negligible in general because of small Wilson 
 coefficients, and  $b_4$ is also small due to the cancellation between
 $A_1^i$ and $A_2^i$. For the other $b_i$ parameters we have
 \begin{eqnarray}
   b_1&=&\frac{C_F}{N_c^2}C_1 A_1^i=18 \frac{C_F}{N_c^2}C_1 \pi
   \alpha_s (X_A+\frac{\pi^2}{3}-4), \\
   b_2&=&\frac{C_F}{N_c^2}C_2 A_1^i=18 \frac{C_F}{N_c^2}C_2 \pi
   \alpha_s (X_A+\frac{\pi^2}{3}-4), \\
   b_3&\simeq& \frac{C_F}{N_c^2}N_c C_6 A_3^f = \pm 6 \frac{C_F}{N_c^2}
   N_c C_6 r_{\chi} \pi \alpha_s (2 X_A^2-X_A) .
 \end{eqnarray}
 So when $b_1$ or $b_2$ dominates the leading power of the annihilation
 contributions is $X_{A}$. This indicates that in some cases, compared
 with $B \to PP$ decays (where $X_{A}^{2}$ appears), annihilation
 topologies might introduce smaller uncertainties into $B{\to}PV$
 branching ratios.

 In the following we will proceed to analyze some decay modes for
 $B{\to}PV$ in detail.

 \subsection{$B^{0}{\to}{\pi}^{\pm}{\rho}^{\mp}$ decays}
 \label{sec41}

 From Table \ref{tab3} and Table \ref{tab4}, we can see that the
 theoretical estimation of the CP-averaged branching ratio
 ${\cal B}r(B^{0}{\to}{\pi}^{+}{\rho}^{-}+{\pi}^{-}{\rho}^{+})$ is in
 agreement with the measurements of the BaBar and CLEO Collaborations
 within one standard deviation, especially when we consider the
 uncertainties due to the variations of input parameters (see
 Fig.\ref{fig3}).

 The decays of $B^{0}{\to}{\pi}^{\pm}{\rho}^{\mp}$ are $a_{1}$ dominant.
 The ``nonfactorizable'' contributions are small, so there is no distinct
 difference between the results obtained with the QCDF approach and those
 with the NF approach. We can see this point from Table \ref{tab4}. The
 annihilation amplitudes of these decay channels are $b_1$ dominant, so
 they do not contribute large uncertainties to the branching ratios. From
 Fig. \ref{fig3}(c), it is obvious that the hard spectator scattering has
 little impact on the branching ratio. Therefore the main uncertainties in
 the CP-averaged branching ratios of these decay modes originate from the
 form factors and CKM matrix elements (especially $|V_{ub}|$), as can be
 seen from Fig.\ref{fig3}.

 The ratio of branching ratios 
 $\frac{{\cal B}r(B^{0}{\to}{\pi}^{-}{\rho}^{+})}
       {{\cal B}r(B^{0}{\to}{\pi}^{+}{\rho}^{-})}$
 has been discussed in \cite{ali}. From Fig.\ref{fig3}(f), it is obvious
 that ${\cal B}r(B^{0}{\to}{\pi}^{-}{\rho}^{+}) >
       {\cal B}r(B^{0}{\to}{\pi}^{+}{\rho}^{-})$, mainly due to the
 difference between the decay constants: $f_{\rho} > f_{\pi}$. In
 addition, the destructive interference among penguin amplitudes also
 lowers the branching ratio of $B^{0}{\to}{\pi}^{+}{\rho}^{-}$, although
 this effect is relatively small. In fact, because the decay constants of
 vector mesons are generally larger than those of pseudoscalar mesons,
 i.e., $f_V > f_P$, it seems universal that, for the $a_{1}$ dominant
 $B{\to}PV$ decay modes, the decay channels with emitted vector mesons
 have larger branching ratios than the corresponding decay channels with
 emitted pseudoscalar mesons. From Fig.\ref{fig3}(a), it is clear that the
 averaged branching ratio of these decays is only mildly dependent on
 the angle ${\gamma}$, which hints the small contribution from penguin
 amplitudes. So it is reasonable to neglect the penguin amplitudes, which
 leads to
 \begin{equation}
 \frac{{\cal B}r(B^{0}{\to}{\pi}^{-}{\rho}^{+})}
      {{\cal B}r(B^{0}{\to}{\pi}^{+}{\rho}^{-})} {\approx} \Big[
 \frac{f_{\rho}F_{1}^{B{\pi}}(0)}{f_{\pi}A_{0}^{B{\rho}}(0)} \Big]^{2}.
 \label{eq:pi-rho}
 \end{equation}
 Clearly, this ratio is insensitive to the CKM matrix elements, dynamical
 coefficients $a_{i}$, and so on. Since it relates different form factors,
 a measurement of this ratio may be helpful to improve theoretical
 predictability. We draw this ratio versus ${\gamma}$ in
 Fig.\ref{fig3}(f).

 \subsection{$B^{-}{\to}{\pi}{\rho},{\pi}{\omega}$ decays}
 \label{sec42}
 The decays $B^{-}{\to}{\pi}{\rho},{\pi}{\omega}$ are also tree dominant,
 but they are determined by $a_{1}+{\xi}a_{2}$. From Fig.\ref{fig4}(a)
 and Fig.\ref{fig5}(a), we can see that within appropriate ranges of the
 parameters, our results for ${\cal B}r(B^{-}{\to}{\pi}^{-}{\rho}^{0})$
 and ${\cal B}r(B^{-}{\to}{\pi}^{-}{\omega})$ are in good agreement with
 the measurements. For ${\cal B}r(B^{-}{\to}{\pi}^{0}{\rho}^{-})$, the 
 branching ratio is predicted numerically to be large and it should be
 observed readily, although there is only an experimental upper limit so
 far.

 Generally speaking, for the branching ratios, the numerical uncertainties
 due to the variations of the CKM matrix elements and form factors are
 always large. But it is distinctive for
 $B^{-}{\to}{\pi}{\rho},{\pi}{\omega}$ decays that hard spectator
 scattering also causes sizable uncertainties [about $20\%$; see Fig.
 \ref{fig4}(c) and Fig. \ref{fig5}(c)]. This is because, as mentioned 
 above, these decay modes are determined by $a_{1}+{\xi}a_{2}$ and
 ``nonfactorizable'' effects; in particular, terms of hard spectator
 scattering contribute greatly to $a_{2}$. As for annihilation
 contributions, due to the cancellations such as $b_{2}(P,V)-b_{2}(V,P)$
 and $b_{3}(P,V)+b_{3}(V,P)$, $A^{a}(B^{-}{\to}{\pi}^{-}{\rho}^{0})$ and
 $A^{a}(B^{-}{\to}{\pi}^{-}{\omega})$ are dominated by $b_{3}$ and
 $b_{2}$, respectively. So for $B^{-}{\to}{\pi}^{-}{\rho}^{0}$ the
 annihilation topologies also contribute large uncertainties to the
 branching ratio, while for $B^{-}{\to}{\pi}^{-}{\omega}$, the uncertainty
 due to the annihilation parameter $X_A$ is negligible
 [see Fig \ref{fig4}(b) and Fig \ref{fig5}(b)].

 \subsection{$B{\to}K{\phi}$ decays}
 \label{sec43}
 The decays $B^{0}{\to}K^{0}{\phi}$ and $B^{-}{\to}K^{-}{\phi}$ have
 triggered great theoretical interest \cite{0012152,ali,9903453,0008159} 
 because they are pure penguin processes. Experimentally, these two decay
 modes have been observed by the BaBar, Belle and CLEO Collaborations; the
 averaged measurements ignoring correlations are
 ${\cal B}r(B^{0}{\to}K^{0}{\phi})=(7.5{\pm}1.8){\times}10^{-6}$ and
 ${\cal B}r(B^{-}{\to}K^{-}{\phi})=(7.9{\pm}1.2){\times}10^{-6}$. Our
 results within the QCDF framework are in good agreement with the
 experimental data within $1{\sigma}$ [see Fig.\ref{fig6}(a,b)].

 From the expressions for the decay amplitudes in Appendix B of 
 \cite{ali}, we can see that the decays $B{\to}K{\phi}$ only relate to the
 CKM factor $V_{tb}V_{ts}^{\ast}$, which is well determined 
 experimentally. So the decay rates are independent of the angle
 ${\gamma}$ and the uncertainties due to the variations of the CKM
 parameters should be very small.

 For $B{\to}K{\phi}$ decays, it is very interesting that the prediction
 is much different between the generalized factorization (GF) approach
 and the QCDF approach. Within the GF framework, the predicted branching
 ratios of the decays $B{\to}K{\phi}$ are very sensitive to the
 ``nonfactorizable'' effects, they vary from $18{\times}10^{-6}$ to
 $0.4{\times}10^{-6}$ \cite{ali} (or from $13{\times}10^{-6}$
 to $0.3{\times}10^{-6}$ \cite{9910291}) with the effective number of
 colors $N_{c}^{eff}=2{\sim}{\infty}$. But within the QCDF framework, 
 there is no need to introduce the effective color number $N_{c}^{eff}$.
 However, since $A^{a}(B{\to}K{\phi})$ is dominated by $b_{3}$, the
 branching ratios will be very sensitive to the annihilation parameter
 $X_{A}$. As for hard spectator scattering, since 
 ${\cal A}^{f}(B{\to}K{\phi})$ is dominated by $a_{4}$, the uncertainties
 from the variations of $X_{H}$ should be small. In \cite{0012152}, the
 authors also analyzed the decays $B{\to}K{\phi}$ with the QCDF approach.
 The effects of the twist-3 LCDAs of $K$ mesons on the hard spectator
 interactions were considered there in spite of their small corrections to
 the decay amplitudes. When the contributions from the weak annihilations
 are taken into account, their predictions are ${\cal B}r(B^{0}{\to}K^{0}
 {\phi})=(4.0_{-1.4}^{+2.9}){\times}10^{-6}$ and ${\cal B}r(B^{-}{\to}
 K^{-}{\phi})=(4.3_{-1.4}^{+3.0}){\times}10^{-6}$ (their central values
 correspond to ${\varrho}_{H}={\varrho}_{A}=0$), which are consistent with
 our estimation at ${\mu}=m_{b}$, although the values of input parameters
 are slightly different.

 \subsection{$B{\to}{\eta}K^{\ast}$ decays}
 \label{sec44}
 The $B{\to}{\eta}K^{\ast}$ decays are well established experimentally by
 the BaBar, Belle, and CLEO Collaborations (see Table \ref{tab3}). But
 within the QCDF framework our estimations (see Table \ref{tab5}) using
 default input parameters are several times smaller than the experimental
 measurements. Even considering the large numerical uncertainties, the
 predicted branching ratios are only marginally consistent with the
 experiments [see Fig. \ref{fig6}(c,d)].

 In fact, there is a similar problem for $B{\to}{\eta}^{(\prime)}K$
 decays which aroused intense discussion several years ago (see, for
 example, Refs. \cite{9704412,9707251,9712372,9711428,9710509,0012208}). 
 It is now commonly believed that this puzzle may be related to a special
 property of ${\eta}^{\prime}$: it has large coupling with two gluons
 \cite{9711428,9710509}. However, there are definitely no large ${\eta}gg$
 contributions to the decay amplitudes of $B{\to}{\eta}K^{\ast}$.

 Here we would like to point out that the CP-averaged branching ratios for
 $B{\to}{\eta}^{\prime}K^{\ast}$ decays are similar to those for
 $B{\to}{\eta}K^{\ast}$ decays with the default parameters (see Table
 \ref{tab5}), which are different from the results listed in
 \cite{9910291,ali,0012208}. H. Cheng and K. Yang thought that
 ${\cal B}r(B{\to}{\eta}K^{\ast})$ should be much larger than
 ${\cal B}r(B{\to}{\eta}^{\prime}K^{\ast})$ in the GF approach, and
 their favorable values are
 $\frac{{\cal B}r(B^{0}{\to}{\eta}K^{{\ast}0})}
 {{\cal B}r(B^{0}{\to}{\eta}^{\prime}K^{{\ast}0})}{\simeq}20$ and
 $\frac{{\cal B}(B^{-}{\to}{\eta}K^{{\ast}-})}
 {{\cal B}r(B^{-}{\to}{\eta}^{\prime}K^{{\ast}-})}{\simeq}18$ with
 $N_{c}^{eff}(LL)=2$ and $N_{c}^{eff}(LR)=6$ \cite{9910291}. In 
 \cite{ali}, A. Ali {\em et al.} predicted that 
 ${\cal B}r(B{\to}{\eta}^{\prime}K^{\ast})$ would be too small to exceed
 $1{\times}10^{-6}$. In those works \cite{9910291,ali,0012208}, the small
 branching ratios for $B{\to}{\eta}^{\prime}K^{\ast}$ decays are due to
 delicate cancellation between different parts of the decay amplitudes.
 However, this cancellation is sensitive to the choice of the input
 parameters such as form factors and so on (see Table. IV in 
 \cite{9910291}). In our analyses, we find that the weak annihilation
 contributions might cause large uncertainties in the predictions of
 $B{\to}{\eta}^{(\prime)}K^{\ast}$ decays.

 Let us have a closer look at the numerical uncertainties of
 ${\cal B}r(B{\to}{\eta}K^{\ast})$. Their decay amplitudes are penguin
 dominant, so for the CKM factors only $|V_{tb}V_{ts}^{\ast}|$ plays an
 important role. Since $|V_{tb}V_{ts}^{\ast}|$ is well determined 
 experimentally, we do not expect large uncertainties from the CKM
 parameters. As for hard spectator scattering, since $a_{4}$ and $a_{6}$
 are the most important coefficients in the amplitudes, the hard spectator
 parameter $X_{H}$ has a small effect on the branching ratios. The decay
 amplitudes also depend on the form factors $F_{1}^{B{\to}{\eta}}$ and
 $A_{0}^{B{\to}K^{\ast}}$, but it seems unlikely that the form factors
 would be dramatically different from the default values. Fortunately, the
 annihilation contribution is dominated by $b_{3}$, which could bring large 
 uncertainties to the branching ratios. However, considering the big gap
 between experiments and default numerical predictions, the annihilation
 parameter $X_{A}$ should be very large. In this paper, we choose the
 model that $X_{A}$ is universal for all $B{\to}PV$ channels, and even
 universal for all $B{\to}PP$ channels \cite{0108141}, so is it possible
 for a large $X_{A}$ to survive for a global $B{\to}PV$ and $PP$ fit?
 It may be a challenge for the QCDF approach to obtain a large
 ${\cal B}r(B{\to}{\eta}K^{\ast})$.

 \subsection{Other decay modes}
 \label{sec45}
 There are still some decay modes for $B{\to}PV$ where the QCDF 
 predictions are marginally consistent or even inconsistent with the
 experimental results, such as 
 $B^{-}{\to}{\pi}^{-}{\overline{K}}^{{\ast}0}$ and
 ${\overline{B}}^{0}{\to}K^{-}{\rho}^{+},{\pi}^{+}K^{{\ast}-}$
 (see Fig. \ref{fig6}). As an illustration, let us see what the QCDF
 prediction for $B^{-}{\to}{\pi}^{-}{\overline{K}}^{{\ast}0}$ and its
 uncertainties are. Without considering annihilation topologies, this
 decay channel is a pure penguin process and dominated by the QCD
 coefficient $a_{4}$ and the CKM factor $|V_{tb}V_{ts}^{\ast}|$. So the
 branching ratio is independent of the angle ${\gamma}$, and the variation
 of the CKM parameters and hard spectator parameter will have very little
 effect on the decay rate. As for the form factors, only $F^{B{\to}{\pi}}$
 is involved. But from the experience of $B{\to}{\pi}{\pi}$ \cite{0108141}
 we know that it is favorable to have a small form factor. Hence we need
 a large annihilation parameter $X_{A}$ again. This indicates that soft
 interactions play a very important role in some cases. Anyway, it is not
 good news because the soft annihilation part is model dependent. As we
 know, annihilation topologies are very important for $B{\to}PP$ decays;
 therefore it is very important to check whether a large $X_{A}$ could be
 acceptable for other decay modes, such as $B{\to}{\pi}{\pi}$ and 
 ${\pi}K$. Experimentally, it is also important to search for pure
 annihilation processes, such as $B^{0}{\to}K^{+}K^{-}$ and $K^{\pm} 
 K^{{\ast}{\mp}}$, which may be helpful in learning more about the 
 annihilation mechanism.

 Within the QCDF framework, some $B{\to}PV$ decay modes are predicted to
 have small branching ratios. One type is the $a_{2}$ dominant neutral B
 decays, such as ${\overline{B}}^{0}{\to}{\pi}^{0}{\rho}^{0},{\pi}^{0}{\omega},
 {\eta}^{(\prime)}{\rho}^{0},{\eta}^{(\prime)}{\omega}$ etc. Another type
 is those decays whose tree amplitudes are denoted by $a_{2}$ and 
 suppressed by the CKM factors; moreover, the interference among penguin
 amplitudes is destructive, for example, the decays
 ${\overline{B}}^{0}{\to}{\overline{K}}^{0}{\rho}^{0},
 {\overline{K}}^{0}{\omega}$ can be classified into this type. The third
 type is pure penguin processes including
 $B^{-}{\to}{\overline{K}}^{0}{\rho}^{-},K^{0}K^{{\ast}-}$ and
 ${\overline{B}}^{0}{\to}K^{0}{\overline{K}}^{{\ast}0}$; their small 
 branching ratios are due to the delicate cancellations among various 
 competing terms. The last type is those decays that are governed by the
 small coefficients $a_{i}$, or electroweak penguin dominant decays, 
 including $B^{-}{\to}{\phi}{\pi}^{-}$ and
 ${\overline{B}}^{0}{\to}{\phi}{\pi}^{0},{\phi}{\eta}^{(\prime)}$, etc.
 Our numerical results in Table \ref{tab4} and Table \ref{tab5} indicate
 that the CP-averaged branching ratios of the above-mentioned decay modes
 are indeed small, and do not exceed $1{\times}10^{-6}$ without 
 considering the uncertainties. They are much below the corresponding
 experimental upper limits. But for these decays, the effects of weak
 annihilations, soft final-state interactions, and other kinds of power
 corrections might be very important or even dominant.

 \section{CP violation}
 \label{sec5}
 CP violation and quark mixing are closely related to each other in the 
 SM. The existence of a CP-violating phase in the top sector of the CKM
 matrix has been established, i.e., $\text{Im}(V_{td}){\propto}\bar{\eta}
 {\neq}0$. Large CP-violating effects are anticipated and have been
 observed in the B meson system. In this section, we will study CP
 violation for charmless decays of $B{\to}PV$ within the QCDF approach.

 Suppose that the decay amplitude for $B{\to}f$ can be expressed as
 \begin{equation}
  {\cal A}(B{\to}f)=A_{1}e^{-i{\xi}_{1}}e^{-i{\delta}_{1}}
                   +A_{2}e^{-i{\xi}_{2}}e^{-i{\delta}_{2}},
 \label{eq:cp-1}
 \end{equation}
 where the $A_{i}$ are the magnitudes, and ${\xi}_{i}$ and ${\delta}_{i}$
 are the strong and weak phases, respectively. Then we can get the
 CP-violating asymmetries
  \begin{eqnarray}
  {\cal A}_{CP}&=&
            \frac{{\Gamma}(B{\to}f)-{\Gamma}(\overline{B}{\to}\bar{f})}
                 {{\Gamma}(B{\to}f)+{\Gamma}(\overline{B}{\to}\bar{f})}
          = \frac{{\vert}{\cal A}(B{\to}f){\vert}^{2}-
                  {\vert}{\cal A}(\overline{B}{\to}\bar{f}){\vert}^{2}}
                 {{\vert}{\cal A}(B{\to}f){\vert}^{2}+
                  {\vert}{\cal A}(\overline{B}{\to}\bar{f}){\vert}^{2}}
         \nonumber \\ &=&
            \frac{2A_{1}A_{2}{\sin}({\xi}_{1}-{\xi}_{2})
           {\sin}({\delta}_{1}-{\delta}_{2})}
           {A_{1}^{2}+A_{2}^{2}+2A_{1}A_{2}{\cos}({\xi}_{1}-{\xi}_{2})
           {\cos}({\delta}_{1}-{\delta}_{2})},
  \label{eq:cp-2}
  \end{eqnarray}
 clearly, ${\cal A}_{CP}{\propto}{\sin}({\xi}_{1}-{\xi}_{2})$. So we would
 like to repeat some remarks. (1) Within the NF framework, there are no
 direct CP violations for the hadronic charmless two-body B decays, 
 because in the decay amplitudes the Wilson coefficients and the hadronic
 matrix elements which are expressed by the product of the decay 
 constants and form factors, are all real, so the strong phase shift
 ${\xi}_{1}-{\xi}_{2}=0$. This indicates that ``nonfactorizable'' effects
 are important for CP violations in hadronic B decays. (2) In the QCDF 
 framework, the CP violations for the class I ($a_{1}$ dominant) decay
 modes should be small, because the strong phases arise at the order of
 ${\alpha}_{s}$ and/or ${\Lambda}_{QCD}/m_{b}$, and hence 
 ${\sin}({\xi}_{1}-{\xi}_{2})$ is small in general. For class II ($a_{2}$
 dominant) decay modes, since $a_{2,I}$ has a large imaginary part (see
 Table \ref{tab2}), a large CP violation might occur.

 For charged B decays, the final states are self-tagging. The
 direct CP-violating asymmetries are defined as
 \begin{equation}
 {\cal A}_{CP}=
       \frac{{\Gamma}(B^{-}{\to}f^{-})-{\Gamma}(B^{+}{\to}f^{+})}
            {{\Gamma}(B^{-}{\to}f^{-})-{\Gamma}(B^{+}{\to}f^{+})}.
 \label{eq:cp-charged}
 \end{equation}

 For neutral B mesons, effects of $B^0$-${\overline{B}}^{0}$ oscillation
 will induce CP asymmetries, so time-dependent measurements of 
 CP-violating asymmetries are needed:
 \begin{equation}
 {\cal A}_{CP}(t)=\frac{{\Gamma}({\overline{B}}^{0}(t){\to}\bar{f})
                       -{\Gamma}(B^{0}(t){\to}f)}
                       {{\Gamma}({\overline{B}}^{0}(t){\to}\bar{f})
                       +{\Gamma}(B^{0}(t){\to}f)}.
 \label{eq:cp-neutral-1}
 \end{equation}
 As discussed in \cite{ali,9501295}, here we consider three cases for
 decays of ${\overline{B}}^{0}$ and $B^{0}{\to}PV$.
 \begin{itemize}
 \item case 1: $B^{0}{\to}f$, ${\overline{B}}^{0}{\to}\bar{f}$, but
       $B^{0}{\not\to}\bar{f}$, ${\overline{B}}^{0}{\not\to}f$, for
       example, ${\overline{B}}^{0}{\to}K^{-}{\rho}^{+}$. CP-violating
       asymmetries for these decays are similar to those for $B^{\pm}$
       decays, and no mixing is involved.
 \item case 2: $B^{0}{\to}(f=\bar{f}){\leftarrow}{\overline{B}}^{0}$,
       for example, ${\overline{B}}^{0}{\&}B^{0}{\to}{\pi}^{0}{\rho}^{0},
       {\eta}{\omega}$, etc. The time-integrated asymmetries are
      \begin{equation}
       {\cal A}_{CP}=
          \frac{1}{1+x^{2}}a_{{\epsilon}^{\prime}}
         +\frac{x}{1+x^{2}}a_{{\epsilon}+{\epsilon}^{\prime}},
       \label{eq:cp-neutral-2b}
       \end{equation}
       \begin{equation}
       a_{{\epsilon}^{\prime}}=
           \frac{1-{\vert}{\lambda}_{CP}{\vert}^{2}}
                {1+{\vert}{\lambda}_{CP}{\vert}^{2}},\ \ \
       a_{{\epsilon}+{\epsilon}^{\prime}}=
           \frac{-2\text{ Im}({\lambda}_{CP})}
                {1+{\vert}{\lambda}_{CP}{\vert}^{2}},\ \ \
       {\lambda}_{CP}=
          \frac{V_{td}V_{tb}^{\ast}}{V_{td}^{\ast}V_{tb}}\;
          \frac{{\cal A}({\overline{B}}^{0}(0){\to}\bar{f})}
               {{\cal A}(B^{0}(0){\to}f)},
       \label{eq:cp-neutral-2a}
       \end{equation}

       where $x={\Delta}m/{\Gamma}=0.73{\pm}0.03$ \cite{PDG2000}.
       $a_{{\epsilon}^{\prime}}$ and $a_{{\epsilon}+{\epsilon}^{\prime}}$
       are direct and mixing-induced CP-violating asymmetries, respectively.
 \item case 3: $B^{0}{\rightarrow}(f$ and $(\bar{f}){\leftarrow}
       {\overline{B}}^{0}$. There are three examples for $B{\to}PV$
       decays; they are ${\overline{B}}^{0}$ and
       $B^{0}{\to}{\pi}^{\pm}{\rho}^{\mp},K^{\pm}K^{{\ast}{\mp}}$ and
       ${\overline{B}}^{0}$ and $B^{0}{\to}K^{0}_{S}K^{{\ast}0},
       K^{0}_{S}{\overline{K}}^{{\ast}0}$. We follow the conventions for
       the time-dependent asymmetries in \cite{ali,9501295}. The four
       basic decay amplitudes of transitions ${\overline{B}}^{0}(t)$ 
       and $B^{0}(t){\to}f$ and $\bar{f}$ at $t=0$ are written as
       \begin{mathletters}
       \begin{eqnarray}
          & &g={\cal A}(B^{0}(0){\to}f),\ \ \ \ \ \ \ \ \ \
       \bar{g}={\cal A}({\overline{B}}^{0}(0){\to}\bar{f}),\\
          & &h={\cal A}({\overline{B}}^{0}(0){\to}f),\ \ \ \ \ \ \ \ \ \
       \bar{h}={\cal A}(B^{0}(0){\to}\bar{f}),
       \end{eqnarray}
       \label{eq:cp-neutral-3a}
       \end{mathletters}
       For example, if $f=K^{+}K^{{\ast}-}$, then $h$ and $\bar{g}$ will
       correspond to Eq.(\ref{eq:appendix-2}) and Eq.(\ref{eq:appendix-6}).
       The time-dependent decay widths are written as
       \begin{mathletters}
       \begin{eqnarray}
       & &{\Gamma}(B^{0}(t){\to}f)=
           \frac{e^{-{\Gamma}t}}{2}
           \Big( {\vert}g{\vert}^{2}+{\vert}h{\vert}^{2} \Big)
           \Big[1+a_{{\epsilon}^{\prime}}{\cos}({\Delta}mt)
                 +a_{{\epsilon}+{\epsilon}^{\prime}}
                     {\sin}({\Delta}mt) \Big],\\
       & &{\Gamma}({\overline{B}}^{0}(t){\to}\bar{f})=
           \frac{e^{-{\Gamma}t}}{2}
           \Big( {\vert}\bar{g}{\vert}^{2}+{\vert}\bar{h}{\vert}^{2} \Big)
           \Big[1-a_{{\overline{\epsilon}}^{\prime}}{\cos}({\Delta}mt)
                 -a_{{\epsilon}+{\overline{\epsilon}}^{\prime}}
                     {\sin}({\Delta}mt) \Big],\\
       & &{\Gamma}(B^{0}(t){\to}\bar{f})=
           \frac{e^{-{\Gamma}t}}{2}
           \Big( {\vert}\bar{g}{\vert}^{2}+{\vert}\bar{h}{\vert}^{2} \Big)
           \Big[1+a_{{\overline{\epsilon}}^{\prime}}{\cos}({\Delta}mt)
                 +a_{{\epsilon}+{\overline{\epsilon}}^{\prime}}
                     {\sin}({\Delta}mt) \Big],\\
       & &{\Gamma}({\overline{B}}^{0}(t){\to}f)=
           \frac{e^{-{\Gamma}t}}{2}
           \Big( {\vert}g{\vert}^{2}+{\vert}h{\vert}^{2} \Big)
           \Big[1-a_{{\epsilon}^{\prime}}{\cos}({\Delta}mt)
                 -a_{{\epsilon}+{\epsilon}^{\prime}}
                     {\sin}({\Delta}mt) \Big],
       \end{eqnarray}
       \label{eq:cp-neutral-3c}
       \end{mathletters}
       where $q/p=\frac{V_{td}V_{tb}^{\ast}}{V_{td}^{\ast}V_{tb}}$, and
       \begin{mathletters}
       \begin{eqnarray}
       & &a_{{\epsilon}^{\prime}}
          =\frac{{\vert}g{\vert}^{2}-{\vert}h{\vert}^{2}}
                {{\vert}g{\vert}^{2}+{\vert}h{\vert}^{2}},
              \ \ \ \ \ \
          a_{{\epsilon}+{\epsilon}^{\prime}}
          =\frac{-2\text{ Im}[(q/p){\times}(h/g)]}
                {1+{\vert}h/g{\vert}^{2}},\\
       & &a_{{\overline{\epsilon}}^{\prime}}
          =\frac{{\vert}\bar{h}{\vert}^{2}-{\vert}\bar{g}{\vert}^{2}}
                {{\vert}\bar{h}{\vert}^{2}+{\vert}\bar{g}{\vert}^{2}},
              \ \ \ \ \ \
          a_{{\epsilon}+{\overline{\epsilon}}^{\prime}}
          =\frac{-2\text{ Im}[(q/p){\times}({\bar{g}}/{\bar{h}})]}
                {1+{\vert}{\bar{g}}/{\bar{h}}{\vert}^{2}}.
       \end{eqnarray}
       \label{eq:cp-neutral-3b}
       \end{mathletters}
 \end{itemize}

 Our results for CP-violating asymmetries for decays $B{\to}PV$ are listed
 in Tables \ref{tab6}--\ref{tab8}. The parameters (including
 $a_{{\epsilon}^{\prime}}$, $a_{{\overline{\epsilon}}^{\prime}}$,
 $a_{{\epsilon}+{\epsilon}^{\prime}}$,
 $a_{{\epsilon}+{\overline{\epsilon}}^{\prime}}$, and ${\cal A}_{CP}$)
 with superscripts $f$ and $f+a$ are calculated in the QCDF framework  
 with decay amplitudes ${\cal A}^{f}(B{\to}PV)$ and 
 ${\cal A}^{f}(B{\to}PV)+{\cal A}^{a}(B{\to}PV)$, respectively.
 Some remarks are in order.

 %\subsection{CP violation}
 %\label{sec51}

 (1) Within the QCDF framework, the strong phases are either at the
     order of ${\alpha}_{s}$ or power suppressed in
     ${\Lambda}_{QCD}/m_{b}$. Only radiative corrections are in principle
     calculable in the QCDF approach. However, numerically power
     corrections in ${\Lambda}_{QCD}/m_{b}$ might be as important as the
     radiative corrections. So in a sense the QCDF method can predict
     only the order of magnitude of the CP-violating asymmetries.

 (2) The theoretical predictions for CP-violating asymmetries are 
     compatible with measurements within one standard deviation
    (see Fig. \ref{fig7}). However, so far, the present experimental
     measurements on $B{\to}PV$ have too large uncertainties, e.g.,
     ${\cal A}_{CP}({\omega}{\pi}^{\pm})=-0.34{\pm}0.25$ \cite{0001009}
     or $-0.01^{+0.29}_{-0.31}{\pm}0.03$ \cite{0109006}, and
     ${\cal A}_{CP}({\phi}K^{\pm})=-0.05{\pm}0.20{\pm}0.03$
     \cite{0109006}, so they cannot provide any useful information.

 (3) Although the uncertainties from variations of nonperturbative
     parameters, such as form factors and so on, are reduced to some
     extent for the CP-violating asymmetries, the weak annihilations
     can still have great effects on the CP asymmetries for some decay
     channels,  such as $a_{{\epsilon}^{\prime}}$ and
     $a_{{\epsilon}+{\epsilon}^{\prime}}$ for the $a_{2}$ dominant decays
     ${\overline{B}}^{0}{\to}{\pi}{\omega},{\eta}{\rho}^{0}$, etc.

 (4) It is worth noting that for the decays $B^{0}{\to}{\phi}{\pi},
     {\phi}{\eta}^{(\prime)}$, there are no direct CP asymmetries
     $a_{{\epsilon}^{\prime}}$ within the  QCDF approach, while in the
     GF framework, $a_{{\epsilon}^{\prime}}$ varies from ${\sim}1\%$ to
     ${\sim}23\%$ with $N_{c}^{eff}=2{\sim}{\infty}$ \cite{ali}.

 (5) One might wonder why the CP-violating asymmetries for the $a_{1}$
     dominant decays $B^{0}{\to}{\pi}^{\pm}{\rho}^{\mp}$ are not as small
     as we expected. In fact, we only expect that the direct CP
     asymmetries at $t=0$ are small, i.e.,
     $a_{{\epsilon}^{{\prime}{\prime}}}=
      \frac{{\vert}g{\vert}^{2}-{\vert}\bar{g}{\vert}^{2}}
           {{\vert}g{\vert}^{2}+{\vert}\bar{g}{\vert}^{2}}$ and
     $a_{{\overline{\epsilon}}^{{\prime}{\prime}}}=
      \frac{{\vert}h{\vert}^{2}-{\vert}\bar{h}{\vert}^{2}}
           {{\vert}h{\vert}^{2}+{\vert}\bar{h}{\vert}^{2}}$.
     For instance, we take $g={\cal A}(B^{0}{\to}{\pi}^{-}{\rho}^{+})$;
     then with the CKM parameters in \cite{0112133},
     $a_{{\epsilon}^{{\prime}{\prime}}}^{f}=2.3\%$,
     $a_{{\epsilon}^{{\prime}{\prime}}}^{f+a}=5.2\%$,
     $a_{{\overline{\epsilon}}^{{\prime}{\prime}}}^{f}=-0.2\%$ and
     $a_{{\overline{\epsilon}}^{{\prime}{\prime}}}^{f+a}=4.6\%$
     at the scale ${\mu}=m_{b}$. They are really small.
 
 \section{Conclusion}
 \label{sec6}
 (1) In the heavy quark limit, we calculated the hadronic charmless decays
     $B{\to}PV$ with the QCDF approach including chirally enhanced
     corrections and weak annihilation topologies. Neglecting the power
     suppressed ${\Lambda}_{QCD}/m_{b}$ effects, the ``nonfactorizable''
     contributions that cannot be calculated within the NF framework are
     perturbatively calculable with the QCDF approach at least at the
     order of ${\alpha}_{s}$. They provide the strong phases which are
     important for CP violations.

 (2) Most of the CP-averaged branching ratios are in agreement with
     present measurements, but there are very large theoretical
     uncertainties due to the variations of input parameters. However,
     some decay channels, such as $B{\to}{\eta}K^{\ast}$ and
     $B^{-}{\to}{\pi}^{-}{\overline{K}}^{{\ast}0}$, are only marginally
     consistent with the experimental observations. In these decay
     channels, nonperturbative contributions, such as weak annihilation
     topologies, play a crucial role and need further investigation.

 (3) The CP asymmetries for the $a_{1}$ dominant decays are small
     because of small strong phases, ${\cal O}({\alpha}_{s})$ and/or
     ${\cal O}({\Lambda}_{QCD}/m_{b})$. However, the ``nonfactorizable''
     effects and the weak annihilations can provide large imaginary parts
     for the $a_{2}$ dominant decay amplitudes, and lead to large direct
     CP asymmetries. The theoretical predictions for CP-violating
     asymmetries are compatible with current measurements within one
     standard deviation. It is worth noting that the QCDF predictions
     might give only the proper order of magnitude of the CP asymmetries.

 (4) The present experimental data for nonleptonic B meson decays are
     not yet sufficient, especially for measurements of CP violation. On
     the other hand, there are still many uncertainties in the theoretical
     frame, for instance, the weak annihilations and other potential power
     corrections. Great advances in both experiment and theory in the near
     future are strongly expected.

 \section*{Acknowledgemanets}
 This work was Supported in part by National Natural Science Foundation
 of China. G. Zhu thanks JSPS of Japan for financial support. We thank
 Professor A. Kagan and H. Cheng for their comments on and discussions
 about the manuscript.

 \begin{appendix}
 \section{The annihilation amplitudes for $B{\to}PV$}
 \label{sec:app1}
 \begin{eqnarray}
 {\cal A}^{a}({\overline{B}}^{0}{\to}K^{0}{\overline{K}}^{{\ast}0})
  &=& \frac{G_{F}}{\sqrt{2}} f_{B} f_{K} f_{K^{\ast}}
  \bigg\{ ( V_{ub}V_{ud}^{\ast} + V_{cb}V_{cd}^{\ast} )
  \Big[ b_{3}(K^{0},{\overline{K}}^{{\ast}0})
        \nonumber \\
  & & - \frac{1}{2}b_{3}^{ew}(K^{0},{\overline{K}}^{{\ast}0})
      + b_{4}(K^{0},{\overline{K}}^{{\ast}0})
      - \frac{1}{2}b_{4}^{ew}(K^{0},{\overline{K}}^{{\ast}0})
        \nonumber \\
  & & + b_{4}({\overline{K}}^{{\ast}0},K^{0})
      - \frac{1}{2}b_{4}^{ew}({\overline{K}}^{{\ast}0},K^{0})
  \Big] \bigg\},
 \label{eq:appendix-1}
 \end{eqnarray}
 \begin{eqnarray}
 {\cal A}^{a}({\overline{B}}^{0}{\to}K^{+}K^{{\ast}-})
  &=& \frac{G_{F}}{\sqrt{2}} f_{B} f_{K} f_{K^{\ast}}
  \bigg\{ V_{ub}V_{ud}^{\ast} b_{1}(K^{+},K^{{\ast}-})
        \nonumber \\
  & & + ( V_{ub}V_{ud}^{\ast} + V_{cb}V_{cd}^{\ast} )
  \Big[ b_{4}(K^{+},K^{{\ast}-}) + b_{4}(K^{{\ast}-},K^{+})
        \nonumber \\
  & & + b_{4}^{ew}(K^{+},K^{{\ast}-})
      - \frac{1}{2}b_{4}^{ew}(K^{{\ast}-},K^{+})
  \Big] \bigg\},
 \label{eq:appendix-2}
 \end{eqnarray}
 \begin{eqnarray}
 {\cal A}^{a}({\overline{B}}^{0}{\to}{\pi}^{-}{\rho}^{+})
  &=& \frac{G_{F}}{\sqrt{2}} f_{B} f_{\pi} f_{\rho}
  \bigg\{ V_{ub}V_{ud}^{\ast} b_{1}({\rho}^{+},{\pi}^{-})
        \nonumber \\
  & & + ( V_{ub}V_{ud}^{\ast} + V_{cb}V_{cd}^{\ast} )
  \Big[ b_{3}({\pi}^{-},{\rho}^{+})
      + b_{4}({\rho}^{+},{\pi}^{-})
      + b_{4}({\pi}^{-},{\rho}^{+})
        \nonumber \\
  & & - \frac{1}{2}b_{3}^{ew}({\pi}^{-},{\rho}^{+})
      + b_{4}^{ew}({\rho}^{+},{\pi}^{-})
      - \frac{1}{2}b_{4}^{ew}({\pi}^{-},{\rho}^{+})
  \Big] \bigg\},
 \label{eq:appendix-3}
 \end{eqnarray}
 \begin{eqnarray}
 {\cal A}^{a}({\overline{B}}^{0}{\to}{\pi}^{+}{\rho}^{-})
  & & = \frac{G_{F}}{\sqrt{2}} f_{B} f_{\pi} f_{\rho}
  \bigg\{ V_{ub}V_{ud}^{\ast} b_{1}({\pi}^{+},{\rho}^{-})
        \nonumber \\
  & & + ( V_{ub}V_{ud}^{\ast} + V_{cb}V_{cd}^{\ast} )
  \Big[ b_{3}({\rho}^{-},{\pi}^{+})
      + b_{4}({\pi}^{+},{\rho}^{-})
      + b_{4}({\rho}^{-},{\pi}^{+}) \nonumber \\
  & & - \frac{1}{2}b_{3}^{ew}({\rho}^{-},{\pi}^{+})
      + b_{4}^{ew}({\pi}^{+},{\rho}^{-})
      - \frac{1}{2}b_{4}^{ew}({\rho}^{-},{\pi}^{+})
  \Big] \bigg\},
 \label{eq:appendix-4}
 \end{eqnarray}
 \begin{eqnarray}
 {\cal A}^{a}({\overline{B}}^{0}{\to}{\pi}^{+}K^{{\ast}-})
  &=& \frac{G_{F}}{\sqrt{2}} f_{B} f_{\pi} f_{K^{\ast}}
  \bigg\{ ( V_{ub}V_{us}^{\ast} + V_{cb}V_{cs}^{\ast} )
  \Big[ b_{3}(K^{{\ast}-},{\pi}^{+})
       \nonumber \\
  & & - \frac{1}{2}b_{3}^{ew}(K^{{\ast}-},{\pi}^{+})
  \Big] \bigg\},
 \label{eq:appendix-5}
 \end{eqnarray}
 \begin{eqnarray}
 {\cal A}^{a}({\overline{B}}^{0}{\to}K^{-}K^{{\ast}+})
  &=& \frac{G_{F}}{\sqrt{2}} f_{B} f_{K} f_{K^{\ast}}
  \bigg\{ V_{ub}V_{ud}^{\ast} b_{1}(K^{{\ast}+},K^{-})
       \nonumber \\
  & & + ( V_{ub}V_{ud}^{\ast} + V_{cb}V_{cd}^{\ast} )
  \Big[ b_{4}(K^{{\ast}+},K^{-}) + b_{4}(K^{-},K^{{\ast}+})
       \nonumber \\
  & & + b_{4}^{ew}(K^{{\ast}+},K^{-})
      - \frac{1}{2}b_{4}^{ew}(K^{-},K^{{\ast}+})
  \Big] \bigg\},
 \label{eq:appendix-6}
 \end{eqnarray}
 \begin{eqnarray}
 {\cal A}^{a}({\overline{B}}^{0}{\to}K^{-}{\rho}^{+})
  &=& \frac{G_{F}}{\sqrt{2}} f_{B} f_{K} f_{\rho}
  \bigg\{ ( V_{ub}V_{us}^{\ast} + V_{cb}V_{cs}^{\ast} )
  \Big[ b_{3}(K^{-},{\rho}^{+})
       \nonumber \\
  & & - \frac{1}{2}b_{3}^{ew}(K^{-},{\rho}^{+})
  \Big] \bigg\},
 \label{eq:appendix-7}
 \end{eqnarray}
 \begin{eqnarray}
  {\cal A}^{a}({\overline{B}}^{0}{\to}{\overline{K}}^{0}K^{{\ast}0})
   &=& \frac{G_{F}}{\sqrt{2}} f_{B} f_{K} f_{K^{\ast}}
  \bigg\{ ( V_{ub}V_{ud}^{\ast} + V_{cb}V_{cd}^{\ast} )
  \Big[ b_{3}(K^{{\ast}0},{\overline{K}}^{0})
       \nonumber \\
   & & + b_{4}(K^{{\ast}0},{\overline{K}}^{0})
       + b_{4}({\overline{K}}^{0},K^{{\ast}0})
       - \frac{1}{2}b_{3}^{ew}(K^{{\ast}0},{\overline{K}}^{0})
       \nonumber \\
   & & - \frac{1}{2}b_{4}^{ew}(K^{{\ast}0},{\overline{K}}^{0})
       - \frac{1}{2}b_{4}^{ew}({\overline{K}}^{0},K^{{\ast}0})
  \Big] \bigg\},
 \label{eq:appendix-8}
 \end{eqnarray}
 \begin{eqnarray}
  {\cal A}^{a}({\overline{B}}^{0}{\to}{\overline{K}}^{0}{\phi})
   &=& \frac{G_{F}}{\sqrt{2}} f_{B} f_{K} f_{\phi}
  \bigg\{ ( V_{ub}V_{us}^{\ast} + V_{cb}V_{cs}^{\ast} )
  \Big[ b_{3}({\phi},{\overline{K}}^{0})
      - \frac{1}{2}b_{3}^{ew}({\phi},{\overline{K}}^{0})
  \Big] \bigg\},
 \label{eq:appendix-9}
 \end{eqnarray}
 \begin{eqnarray}
  {\cal A}^{a}({\overline{B}}^{0}{\to}{\overline{K}}^{0}{\rho}^{0})
   &=& - \frac{G_{F}}{2} f_{B} f_{K} f_{\rho}
  \bigg\{ ( V_{ub}V_{us}^{\ast} + V_{cb}V_{cs}^{\ast} )
  \Big[ b_{3}({\overline{K}}^{0},{\rho}^{0})
      - \frac{1}{2}b_{3}^{ew}({\overline{K}}^{0},{\rho}^{0})
  \Big] \bigg\},
 \label{eq:appendix-10}
 \end{eqnarray}
 \begin{eqnarray}
  {\cal A}^{a}({\overline{B}}^{0}{\to}{\overline{K}}^{0}{\omega})
   &=& \frac{G_{F}}{2} f_{B} f_{K} f_{\omega}
  \bigg\{ ( V_{ub}V_{us}^{\ast} + V_{cb}V_{cs}^{\ast} )
  \Big[ b_{3}({\overline{K}}^{0},{\omega})
      - \frac{1}{2}b_{3}^{ew}({\overline{K}}^{0},{\omega})
  \Big] \bigg\},
 \label{eq:appendix-11}
 \end{eqnarray}
 \begin{eqnarray}
  {\cal A}^{a}({\overline{B}}^{0}{\to}{\pi}^{0}{\rho}^{0})
   &=& \frac{G_{F}}{2 \sqrt{2}} f_{B} f_{\pi} f_{\rho}
  \bigg\{ V_{ub}V_{ud}^{\ast} 
  \Big[ b_{1}({\rho}^{0},{\pi}^{0})
      + b_{1}({\pi}^{0},{\rho}^{0}) \Big]
       \nonumber \\
   & & + ( V_{ub}V_{ud}^{\ast} + V_{cb}V_{cd}^{\ast} )
  \Big[ b_{3}({\rho}^{0},{\pi}^{0}) + b_{3}({\pi}^{0},{\rho}^{0})
       \nonumber \\
   & & + 2b_{4}({\pi}^{0},{\rho}^{0}) + 2b_{4}({\rho}^{0},{\pi}^{0})
       - \frac{1}{2}b_{3}^{ew}({\rho}^{0},{\pi}^{0})
       \nonumber \\
   & & - \frac{1}{2}b_{3}^{ew}({\pi}^{0},{\rho}^{0})
       + \frac{1}{2}b_{4}^{ew}({\pi}^{0},{\rho}^{0})
       + \frac{1}{2}b_{4}^{ew}({\rho}^{0},{\pi}^{0})
  \Big] \bigg\},
 \label{eq:appendix-12}
 \end{eqnarray}
 \begin{eqnarray}
  {\cal A}^{a}({\overline{B}}^{0}{\to}{\pi}^{0}{\omega})
   &=& \frac{G_{F}}{2 \sqrt{2}} f_{B} f_{\pi} f_{\omega}
  \bigg\{ V_{ub}V_{ud}^{\ast} 
  \Big[ b_{1}({\omega},{\pi}^{0})
      + b_{1}({\pi}^{0},{\omega}) \Big]
       \nonumber \\
   & & + ( V_{ub}V_{ud}^{\ast} + V_{cb}V_{cd}^{\ast} )
  \Big[ - b_{3}({\pi}^{0},{\omega})
        - b_{3}({\omega},{\pi}^{0})
       \nonumber \\
   & & + \frac{1}{2}b_{3}^{ew}({\pi}^{0},{\omega})
       + \frac{1}{2}b_{3}^{ew}({\omega},{\pi}^{0})
       + \frac{3}{2}b_{4}^{ew}({\pi}^{0},{\omega})
       \nonumber \\
   & & + \frac{3}{2}b_{4}^{ew}({\omega},{\pi}^{0})
  \Big] \bigg\},
 \label{eq:appendix-13}
 \end{eqnarray}
 \begin{eqnarray}
 {\cal A}^{a}({\overline{B}}^{0}{\to}{\pi}^{0}{\phi})&=&0,
 \label{eq:appendix-14}
 \end{eqnarray}
 \begin{eqnarray}
  {\cal A}^{a}({\overline{B}}^{0}{\to}{\pi}^{0}{\overline{K}}^{{\ast}0})
   &=& - \frac{G_{F}}{2} f_{B} f_{\pi} f_{K^{\ast}}
  \bigg\{ ( V_{ub}V_{us}^{\ast} + V_{cb}V_{cs}^{\ast} )
  \Big[ b_{3}({\overline{K}}^{{\ast}0},{\pi}^{0})
       \nonumber \\
   & & - \frac{1}{2}b_{3}^{ew}({\overline{K}}^{{\ast}0},{\pi}^{0})
  \Big] \bigg\},
 \label{eq:appendix-15}
 \end{eqnarray}
 \begin{eqnarray}
  {\cal A}^{a}({\overline{B}}^{0}{\to}{\eta}^{({\prime})}{\rho}^{0})
   &=& \frac{G_{F}}{2} f_{B} f_{{\eta}^{({\prime})}}^{u} f_{\rho}
  \bigg\{ V_{ub}V_{ud}^{\ast} 
  \Big[ b_{1}({\eta}^{({\prime})},{\rho}^{0})
      + b_{1}({\rho}^{0},{\eta}^{({\prime})}) \Big]
      \nonumber \\
   & & + ( V_{ub}V_{ud}^{\ast} + V_{cb}V_{cd}^{\ast} )
  \Big[ - b_{3}({\eta}^{({\prime})},{\rho}^{0})
        - b_{3}({\rho}^{0},{\eta}^{({\prime})})
       \nonumber \\
   & & + \frac{1}{2}b_{3}^{ew}({\eta}^{({\prime})},{\rho}^{0})
       + \frac{1}{2}b_{3}^{ew}({\rho}^{0},{\eta}^{({\prime})})
       + \frac{3}{2}b_{4}^{ew}({\eta}^{({\prime})},{\rho}^{0})
       \nonumber \\
   & & + \frac{3}{2}b_{4}^{ew}({\rho}^{0},{\eta}^{({\prime})})
  \Big] \bigg\},
 \label{eq:appendix-16}
 \end{eqnarray}
 \begin{eqnarray}
  {\cal A}^{a}({\overline{B}}^{0}{\to}{\eta}^{({\prime})}{\omega})
   &=& \frac{G_{F}}{2} f_{B} f_{{\eta}^{({\prime})}}^{u} f_{\omega}
  \bigg\{ V_{ub}V_{ud}^{\ast}
  \Big[ b_{1}({\eta}^{({\prime})},{\omega})
      + b_{1}({\omega},{\eta}^{({\prime})}) \Big]
       \nonumber \\
   & & + ( V_{ub}V_{ud}^{\ast} + V_{cb}V_{cd}^{\ast} )
  \Big[ b_{3}({\eta}^{({\prime})},{\omega})
       + b_{3}({\omega},{\eta}^{({\prime})})
       \nonumber \\
   & & + 2b_{4}({\eta}^{({\prime})},{\omega})
       + 2b_{4}({\omega},{\eta}^{({\prime})})
       - \frac{1}{2}b_{3}^{ew}({\eta}^{({\prime})},{\omega})
       \nonumber \\
   & & - \frac{1}{2}b_{3}^{ew}({\omega},{\eta}^{({\prime})})
       + \frac{1}{2}b_{4}^{ew}({\eta}^{({\prime})},{\omega})
       + \frac{1}{2}b_{4}^{ew}({\omega},{\eta}^{({\prime})})
  \Big] \bigg\},
 \label{eq:appendix-17}
 \end{eqnarray}
 \begin{eqnarray}
  {\cal A}^{a}({\overline{B}}^{0}{\to}{\eta}^{({\prime})}{\phi})
   &=& \frac{G_{F}}{\sqrt{2}} f_{B} f_{{\eta}^{({\prime})}}^{s} f_{\phi}
  \bigg\{ ( V_{ub}V_{ud}^{\ast} + V_{cb}V_{cd}^{\ast} )
  \Big[ b_{4}({\eta}^{({\prime})},{\phi})
        \nonumber \\
   & & + b_{4}({\phi},{\eta}^{({\prime})})
       - \frac{1}{2}b_{4}^{ew}({\eta}^{({\prime})},{\phi})
       - \frac{1}{2}b_{4}^{ew}({\phi},{\eta}^{({\prime})})
  \Big] \bigg\},
 \label{eq:appendix-18}
 \end{eqnarray}
 \begin{eqnarray}
{\cal A}^{a}({\overline{B}}^{0}{\to}{\eta}^{({\prime})}{\overline{K}}^{{\ast}0})
   & & = \frac{G_{F}}{\sqrt{2}} f_{B} f_{{\eta}^{({\prime})}}^{u} f_{K^{\ast}}
  \bigg\{ ( V_{ub}V_{us}^{\ast} + V_{cb}V_{cs}^{\ast} )
  \Big[ b_{3}({\overline{K}}^{{\ast}0},{\eta}^{({\prime})})
        \nonumber \\
   & & - \frac{1}{2}b_{3}^{ew}({\overline{K}}^{{\ast}0},{\eta}^{({\prime})})
       + \frac{f_{{\eta}^{({\prime})}}^{s}}{f_{{\eta}^{({\prime})}}^{u}}
  \Big( b_{3}({\eta}^{({\prime})},{\overline{K}}^{{\ast}0})
      - \frac{1}{2}b_{3}^{ew}({\eta}^{({\prime})},{\overline{K}}^{{\ast}0}) \Big)
  \Big] \bigg\},
 \label{eq:appendix-19}
 \end{eqnarray}
 \begin{eqnarray}
  {\cal A}^{a}(B^{-}{\to}K^{0}K^{{\ast}-})
   & & = \frac{G_{F}}{\sqrt{2}} f_{B} f_{K} f_{K^{\ast}}
  \bigg\{ V_{ub}V_{ud}^{\ast} b_{2}(K^{0},K^{{\ast}-})
       \nonumber \\
   & & + ( V_{ub}V_{ud}^{\ast} + V_{cb}V_{cd}^{\ast} )
  \Big[ b_{3}(K^{0},K^{{\ast}-}) + b_{3}^{ew}(K^{0},K^{{\ast}-})
  \Big] \bigg\},
 \label{eq:appendix-20}
 \end{eqnarray}
 \begin{eqnarray}
  {\cal A}^{a}(B^{-}{\to}{\pi}^{-}{\rho}^{0})
   &=& \frac{G_{F}}{2} f_{B} f_{\pi} f_{\rho}
  \bigg\{ V_{ub}V_{ud}^{\ast}
  \Big[ b_{2}({\pi}^{-},{\rho}^{0}) - b_{2}({\rho}^{0},{\pi}^{-}) \Big]
       \nonumber \\
   & & + ( V_{ub}V_{ud}^{\ast} + V_{cb}V_{cd}^{\ast} )
  \Big[ b_{3}({\pi}^{-},{\rho}^{0}) - b_{3}({\rho}^{0},{\pi}^{-})
       \nonumber \\
   & & + b_{3}^{ew}({\pi}^{-},{\rho}^{0})
       - b_{3}^{ew}({\rho}^{0},{\pi}^{-})
  \Big] \bigg\},
 \label{eq:appendix-21}
 \end{eqnarray}
 \begin{eqnarray}
  {\cal A}^{a}(B^{-}{\to}{\pi}^{-}{\omega})
   &=& \frac{G_{F}}{2} f_{B} f_{\pi} f_{\omega}
  \bigg\{ V_{ub}V_{ud}^{\ast}
  \Big[ b_{2}({\pi}^{-},{\omega}) + b_{2}({\omega},{\pi}^{-}) \Big]
        \nonumber \\
   & & + ( V_{ub}V_{ud}^{\ast} + V_{cb}V_{cd}^{\ast} )
  \Big[ b_{3}({\pi}^{-},{\omega}) + b_{3}({\omega},{\pi}^{-})
        \nonumber \\
   & & + b_{3}^{ew}({\pi}^{-},{\omega})
       + b_{3}^{ew}({\omega},{\pi}^{-})
  \Big] \bigg\}, 
 \label{eq:appendix-22}
 \end{eqnarray}
 \begin{eqnarray}
  {\cal A}^{a}(B^{-}{\to}{\pi}^{-}{\phi})&=&0,
 \label{eq:appendix-23}
 \end{eqnarray}
 \begin{eqnarray}
  {\cal A}^{a}(B^{-}{\to}{\pi}^{-}{\overline{K}}^{{\ast}0})
   &=& \frac{G_{F}}{\sqrt{2}} f_{B} f_{\pi} f_{K^{\ast}}
  \bigg\{ V_{ub}V_{us}^{\ast}b_{2}({\overline{K}}^{{\ast}0},{\pi}^{-})
        \nonumber \\
   & & + ( V_{ub}V_{us}^{\ast} + V_{cb}V_{cs}^{\ast} )
  \Big[ b_{3}({\overline{K}}^{{\ast}0},{\pi}^{-})
      + b_{3}^{ew}({\overline{K}}^{{\ast}0},{\pi}^{-})
  \Big] \bigg\},
 \label{eq:appendix-24}
 \end{eqnarray}
 \begin{eqnarray}
  {\cal A}^{a}(B^{-}{\to}{\pi}^{0}{\rho}^{-})
   &=& \frac{G_{F}}{2} f_{B} f_{\pi} f_{\rho}
  \bigg\{ V_{ub}V_{ud}^{\ast}
  \Big[ b_{2}({\rho}^{-},{\pi}^{0})
      - b_{2}({\pi}^{0},{\rho}^{-}) \Big]
       \nonumber \\
   & & + ( V_{ub}V_{ud}^{\ast} + V_{cb}V_{cd}^{\ast} )
  \Big[ b_{3}({\rho}^{-},{\pi}^{0}) - b_{3}({\pi}^{0},{\rho}^{-})
        \nonumber \\
   & & + b_{3}^{ew}({\rho}^{-},{\pi}^{0})
       - b_{3}^{ew}({\pi}^{0},{\rho}^{-})
  \Big] \bigg\},
 \label{eq:appendix-25}
 \end{eqnarray}
 \begin{eqnarray}
  {\cal A}^{a}(B^{-}{\to}{\pi}^{0}K^{{\ast}-})
   &=& \frac{G_{F}}{2} f_{B} f_{\pi} f_{K^{\ast}}
  \bigg\{ V_{ub}V_{us}^{\ast} b_{2}(K^{{\ast}-},{\pi}^{0})
       \nonumber \\
   & & + ( V_{ub}V_{us}^{\ast} + V_{cb}V_{cs}^{\ast} )
  \Big[ b_{3}(K^{{\ast}-},{\pi}^{0})
      + b_{3}^{ew}(K^{{\ast}-},{\pi}^{0})
  \Big] \bigg\},
 \label{eq:appendix-26}
 \end{eqnarray}
 \begin{eqnarray}
  {\cal A}^{a}(B^{-}{\to}{\eta}^{({\prime})}{\rho}^{-})
   &=& \frac{G_{F}}{\sqrt{2}} f_{B} f_{{\eta}^{({\prime})}}^{u} f_{\rho}
  \bigg\{ V_{ub}V_{ud}^{\ast}
  \Big[ b_{2}({\rho}^{-},{\eta}^{({\prime})})
      + b_{2}({\eta}^{({\prime})},{\rho}^{-}) \Big]
        \nonumber \\
   & & + ( V_{ub}V_{ud}^{\ast} + V_{cb}V_{cd}^{\ast} )
  \Big[ b_{3}({\rho}^{-},{\eta}^{({\prime})})
      + b_{3}({\eta}^{({\prime})},{\rho}^{-})
      \nonumber \\
   & & + b_{3}^{ew}({\rho}^{-},{\eta}^{({\prime})})
      + b_{3}^{ew}({\eta}^{({\prime})},{\rho}^{-})
  \Big] \bigg\},
 \label{eq:appendix-27}
 \end{eqnarray}
 \begin{eqnarray}
  {\cal A}^{a}(B^{-}{\to}{\eta}^{({\prime})}K^{{\ast}-})
   &=& \frac{G_{F}}{\sqrt{2}} f_{B} f_{{\eta}^{({\prime})}}^{u} f_{K^{\ast}}
  \bigg\{ V_{ub}V_{us}^{\ast} \Big[ b_{2}(K^{{\ast}-},{\eta}^{({\prime})})
     + \frac{f_{{\eta}^{({\prime})}}^{s}}{f_{{\eta}^{({\prime})}}^{u}}
       b_{2}({\eta}^{({\prime})},K^{{\ast}-}) \Big]
       \nonumber \\
   & & + ( V_{ub}V_{us}^{\ast} + V_{cb}V_{cs}^{\ast} )
  \Big[ b_{3}(K^{{\ast}-},{\eta}^{({\prime})})
      + b_{3}^{ew}(K^{{\ast}-},{\eta}^{({\prime})})
       \nonumber \\
   & & + \frac{f_{{\eta}^{({\prime})}}^{s}}{f_{{\eta}^{({\prime})}}^{u}}
  \Big( b_{3}({\eta}^{({\prime})},K^{{\ast}-})
      + b_{3}^{ew}({\eta}^{({\prime})},K^{{\ast}-}) \Big)
  \Big] \bigg\},
 \label{eq:appendix-28}
 \end{eqnarray}
 \begin{eqnarray}
  {\cal A}^{a}(B^{-}{\to}K^{-}K^{{\ast}0})
   &=& \frac{G_{F}}{\sqrt{2}} f_{B} f_{K} f_{K^{\ast}}
  \bigg\{ V_{ub}V_{ud}^{\ast} b_{2}(K^{{\ast}0},K^{-})
       \nonumber \\
   & & + ( V_{ub}V_{ud}^{\ast} + V_{cb}V_{cd}^{\ast} )
  \Big[ b_{3}(K^{{\ast}0},K^{-}) + b_{3}^{ew}(K^{{\ast}0},K^{-})
  \Big] \bigg\},
 \label{eq:appendix-29}
 \end{eqnarray}
 \begin{eqnarray}
  {\cal A}^{a}(B^{-}{\to}K^{-}{\phi})
   &=& \frac{G_{F}}{\sqrt{2}} f_{B} f_{K} f_{\phi}
  \bigg\{ V_{ub}V_{us}^{\ast} b_{2}({\phi},K^{-})
       \nonumber \\
   & & + ( V_{ub}V_{us}^{\ast} + V_{cb}V_{cs}^{\ast} )
  \Big[ b_{3}({\phi},K^{-}) + b_{3}^{ew}({\phi},K^{-})
  \Big] \bigg\},
 \label{eq:appendix-30}
 \end{eqnarray}
 \begin{eqnarray}
  {\cal A}^{a}(B^{-}{\to}K^{-}{\omega})
   &=& \frac{G_{F}}{2} f_{B} f_{K} f_{\omega}   
  \bigg\{ V_{ub}V_{us}^{\ast} b_{2}(K^{-},{\omega})
        \nonumber \\
   & & + ( V_{ub}V_{us}^{\ast} + V_{cb}V_{cs}^{\ast} )
  \Big[ b_{3}(K^{-},{\omega}) + b_{3}^{ew}(K^{-},{\omega})
  \Big] \bigg\},
 \label{eq:appendix-31}
 \end{eqnarray}
 \begin{eqnarray}
  {\cal A}^{a}(B^{-}{\to}K^{-}{\rho}^{0})
   &=& \frac{G_{F}}{2} f_{B} f_{K} f_{\rho}   
  \bigg\{ V_{ub}V_{us}^{\ast} b_{2}(K^{-},{\rho}^{0})
       \nonumber \\
   & & + ( V_{ub}V_{us}^{\ast} + V_{cb}V_{cs}^{\ast} )
  \Big[ b_{3}(K^{-},{\rho}^{0}) + b_{3}^{ew}(K^{-},{\rho}^{0})
  \Big] \bigg\},
 \label{eq:appendix-32}
 \end{eqnarray}
 \begin{eqnarray}
  {\cal A}^{a}(B^{-}{\to}{\overline{K}}^{0}{\rho}^{-})
   &=& \frac{G_{F}}{\sqrt{2}} f_{B} f_{K} f_{\rho}
  \bigg\{ V_{ub}V_{us}^{\ast} b_{2}({\overline{K}}^{0},{\rho}^{-})
       \nonumber \\
   & & + ( V_{ub}V_{us}^{\ast} + V_{cb}V_{cs}^{\ast} )
  \Big[ b_{3}({\overline{K}}^{0},{\rho}^{-})
      + b_{3}^{ew}({\overline{K}}^{0},{\rho}^{-})
  \Big] \bigg\}.
 \label{eq:appendix-33}
 \end{eqnarray}
 \end{appendix}

 \begin{figure}
 \begin{center}
 \begin{picture}(300,160)(0,30)
 \put(-90,-450){\epsfxsize180mm\epsfbox{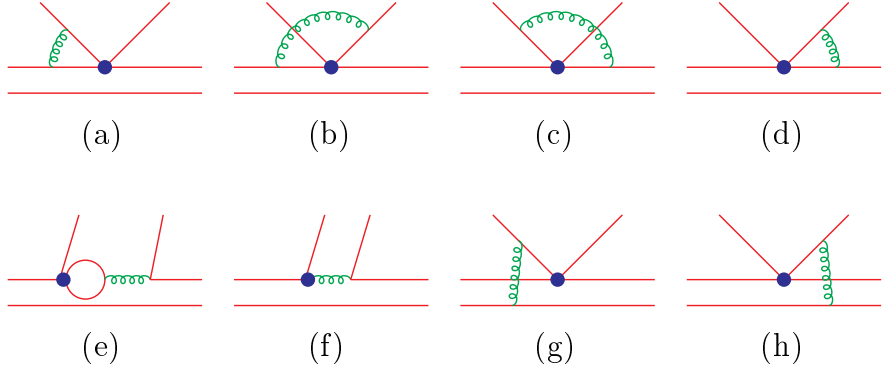}}
 \end{picture}
 \end{center}
 \caption{Order of ${\alpha}_{s}$ corrections to the hard-scattering
      kernels. The upward quark lines represent the emitted mesons from 
      b quark weak decays. These diagrams are commonly called vertex
      corrections, penguin corrections, and hard spectator scattering
      diagrams for Fig.(a-d), Fig.(e-f), and Fig.(g-h) respectively.}
 \label{fig1}
 \end{figure}
 
 \begin{figure}
 \begin{center}
 \begin{picture}(300,100)(0,30) 
 \put(-90,-500){\epsfxsize180mm\epsfbox{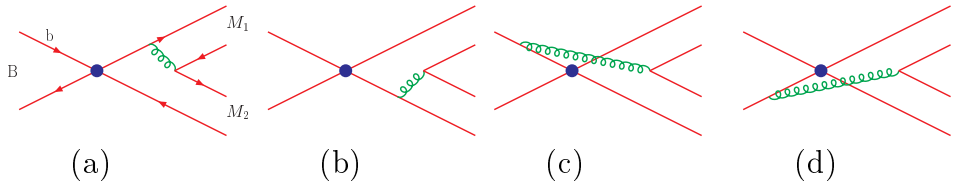}} 
 \end{picture}
 \end{center}
 \caption{Order of ${\alpha}_{s}$ corrections to the weak annihilations
          of charmless decays $B{\to}PV$.}
 \label{fig2}
 \end{figure}

 \begin{figure}
 \begin{center}
 \begin{picture}(300,400)(0,170)
 \put(-115,-100){\epsfxsize190mm\epsfbox{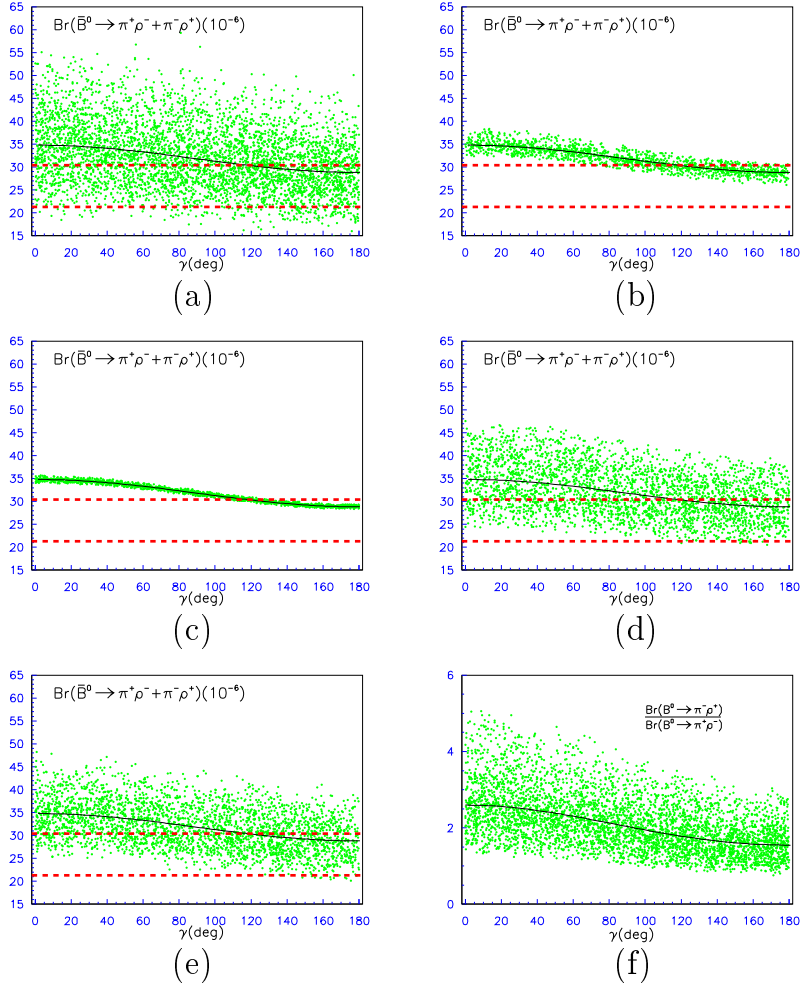}}
 \end{picture}
 \end{center}
 \caption{Decays ${\overline{B}}^{0}$ and
      $B^{0}{\to}{\pi}^{\pm}{\rho}^{\mp}$ versus ${\gamma}$ at the scale
      ${\mu}=m_{b}$ within the  QCDF approach, including the effects of
      weak annihilations. Fig.(a-e) are CP-averaged branching ratios, and
      Fig.(f) is the ratio
      $\frac{{\cal B}r(B^{0}{\to}{\pi}^{-}{\rho}^{+})}
            {{\cal B}r(B^{0}{\to}{\pi}^{+}{\rho}^{-})}$. 
      The solid lines are drawn with central values of various parameters;
      the bands between the dashed lines denote measurement within 
      $1{\sigma}$. The shaded dots originate from uncertainties due to
      variations of parameters, such as in Fig.(b) due to $X_{A}$, in
      Fig.(c) due to $X_{H}$, in Fig.(d) due to form factors, in Fig.(e)
      due to CKM elements, and in Fig.(a,f) due to various parameters.}
 \label{fig3}
 \end{figure}

 \begin{figure}
 \begin{center}
 \begin{picture}(300,400)(0,170)
 \put(-115,-100){\epsfxsize190mm\epsfbox{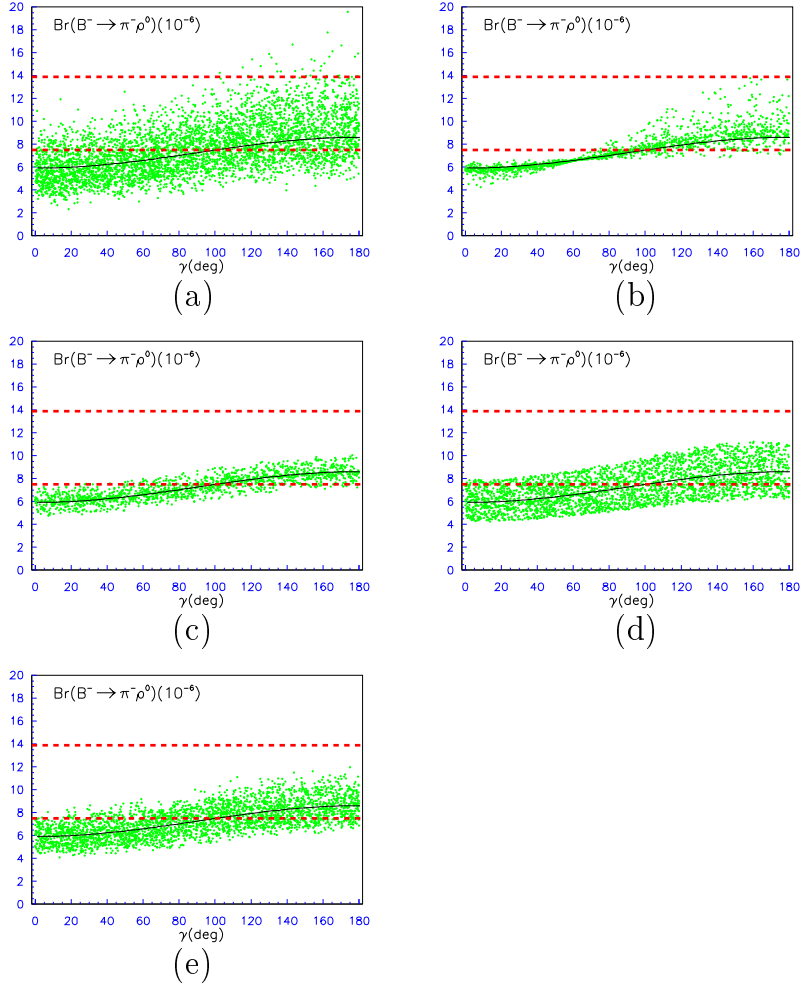}}
 \end{picture}
 \end{center}
 \caption{${\cal B}r(B^{-}{\to}{\pi}^{-}{\rho}^{0})$ versus ${\gamma}$
      at the scale ${\mu}=m_{b}$ with the QCDF approach, including the
      weak annihilation contributions. The solid lines are drawn with
      the default values of various parameters; the bands between the
      dashed lines denote averaged measurement within $1{\sigma}$. The
      shaded dots demonstrate the uncertainties due to variations of
      various parameters in Fig.(a),  $X_{A}$ in Fig.(b), $X_{H}$ in
      Fig.(c), form factors in Fig.(d), and CKM matrix elements in
      Fig.(e).}
 \label{fig4}
 \end{figure}

 \begin{figure}
 \begin{center}
 \begin{picture}(300,400)(0,170)
 \put(-115,-100){\epsfxsize190mm\epsfbox{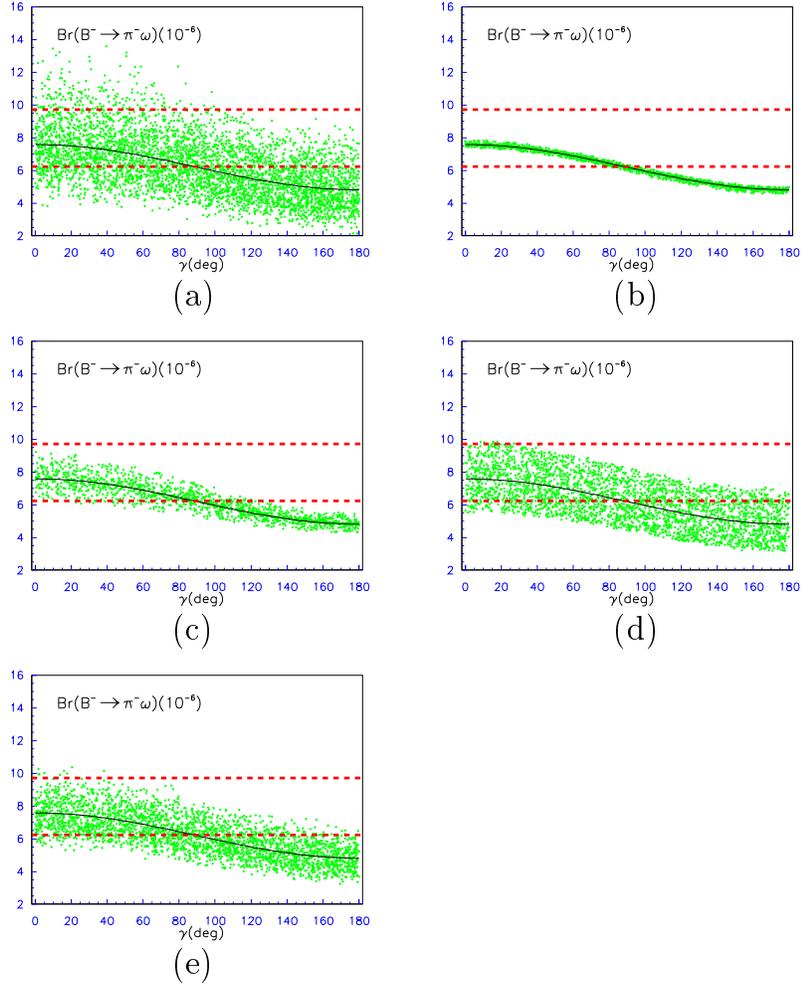}}
 \end{picture}
 \end{center}
 \caption{${\cal B}r(B^{-}{\to}{\pi}^{-}{\omega})$ versus ${\gamma}$
      at the scale ${\mu}=m_{b}$ with the QCDF approach, including the
      weak annihilation contributions. The meaning of the solid lines,
      the bands, and shaded dots is the same as in Fig.\ref{fig4}.}
 \label{fig5}
 \end{figure}

 \begin{figure}
 \begin{center}
 \begin{picture}(300,380)(0,180)
 \put(-115,-100){\epsfxsize190mm\epsfbox{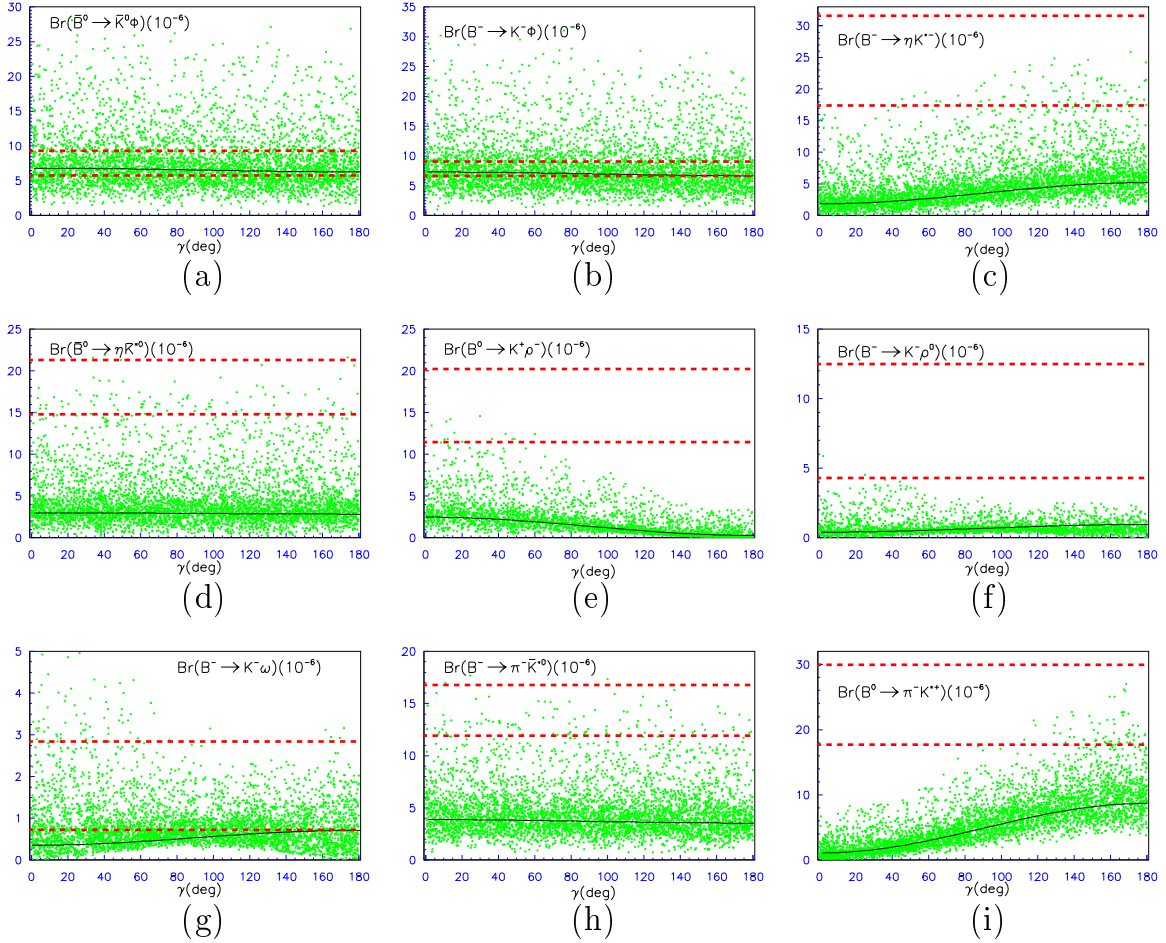}}
 \end{picture}
 \end{center}
 \caption{The CP-averaged branching ratios of $B{\to}PV$ versus ${\gamma}$
      at the scale ${\mu}=m_{b}$ with the QCDF approach, including the
      weak annihilation contributions. The meanings of the solid lines
      and bands are the same as in Fig.\ref{fig4}. The shaded dots denote
      the uncertainties due to variations of various parameters.}
 \label{fig6}
 \end{figure}

 \begin{figure}
 \begin{center}
 \begin{picture}(300,100)(0,550)
 \put(-115,-40){\epsfxsize190mm\epsfbox{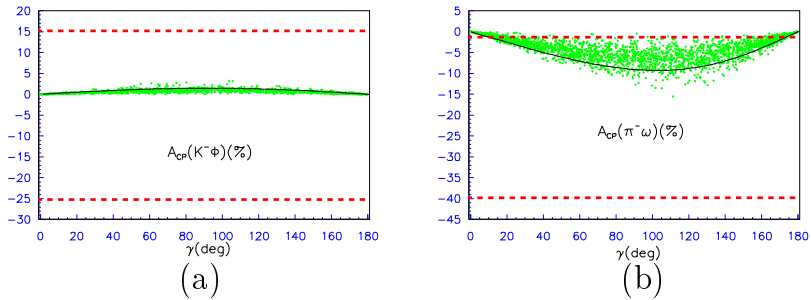}}
 \end{picture}
 \end{center}
 \caption{CP-violating asymmetries ${\cal A}_{CP}(B^{\pm}{\to}K^{\pm}{\phi})$
      in Fig.(a) and ${\cal A}_{CP}(B^{\pm}{\to}{\pi}^{\pm}{\omega})$ in
      Fig.(b) versus ${\gamma}$ at the scale ${\mu}=m_{b}$ with the QCDF
      approach, including the effects of weak annihilations. The meanings 
      of the solid lines, bands, and shaded dots are the same as in
      Fig.\ref{fig4}.}
 \label{fig7}
 \end{figure}

 \begin{table}
 \caption{Wilson coefficients in NDR scheme. The input parameters 
          in numerical calculations are fixed as
          ${\alpha}_{s}(m_{Z})=0.1185$, ${\alpha}_{em}(m_{W})=1/128$, 
          $m_{W}=80.42\text{GeV}$, $m_{Z}=91.188\text{GeV}$, 
          $m_{t}=168.2\text{GeV}$, $m_{b}=4.6\text{GeV}$.}
 \label{tab1}
 \begin{tabular}{ccccccc}
            & \multicolumn{2}{c}{${\mu}=m_{b}/2 $}
            & \multicolumn{2}{c}{${\mu}=m_{b}   $}
            & \multicolumn{2}{c}{${\mu}=2m_{b}  $} \\
              \cline{2-3} \cline{4-5} \cline{6-7}
            & NLO & LO & NLO & LO & NLO & LO \\ \hline
  $C_{1} $  & 1.136 & 1.178 & 1.080 & 1.115 & 1.044 & 1.073 \\
  $C_{2} $  &-0.283 &-0.353 &-0.181 &-0.245 &-0.105 &-0.165 \\
  $C_{3} $  & 0.021 & 0.020 & 0.014 & 0.012 & 0.009 & 0.008 \\
  $C_{4} $  &-0.050 &-0.048 &-0.035 &-0.033 &-0.024 &-0.023 \\
  $C_{5} $  & 0.010 & 0.010 & 0.009 & 0.008 & 0.007 & 0.006 \\
  $C_{6} $  &-0.063 &-0.061 &-0.041 &-0.038 &-0.026 &-0.024 \\
  $C_{7}/{{\alpha}_{em}} $ &-0.020 &-0.103 &-0.004 &-0.096 & 0.019 &-0.081 \\
  $C_{8}/{{\alpha}_{em}} $ & 0.082 & 0.024 & 0.052 & 0.015 & 0.033 & 0.010 \\
  $C_{9}/{{\alpha}_{em}} $ &-1.339 &-0.089 &-1.263 &-0.086 &-1.201 &-0.074 \\
  $C_{10}/{{\alpha}_{em}}$ & 0.369 &-0.022 & 0.253 &-0.016 & 0.168 &-0.012 \\
  $C_{7{\gamma}}         $ &       &-0.341 &       &-0.304 &       &-0.272 \\
  $C_{8g}                $ &       &-0.160 &       &-0.145 &       &-0.132 \\
 \end{tabular}
 \end{table} 
 
 \begin{table}
 \caption{Numerical values of coefficients $a_{i,I}$ with the default
          values of various parameters for case I (the recoiled mesons
          $M_{1}$ are vector mesons, and the emitted mesons $M_{2}$ are
          pseudoscalar mesons).}
 \label{tab2}
 \begin{tabular}{rccc}
   & ${\mu}=m_{b}/2$ & ${\mu}=m_{b}$ & ${\mu}=2m_{b}$ \\ \hline
  $a_{1,I}\ \ \ \ $ 
      &  1.077+0.033i & 1.055+0.018i & 1.038+0.010i \\
  $a_{2,I}\ \ \ \ $ 
      & -0.023-0.110i & 0.018-0.082i & 0.057-0.065i \\ \hline
  $a_{3,I}(10^{-4})$
      & 91.649+44.722i & 70.951+24.259i & 49.938+14.036i \\
  $a_{4,I}^{u}(10^{-4})$
      & -331.4-188.31i &-299.26-152.67i &-270.97-129.36i \\
  $a_{4,I}^{c}(10^{-4})$
      &-406.48-60.422i &-355.33-57.162i &-315.76-53.065i \\
  $a_{5,I}(10^{-4})$
      &-99.038-56.834i &-65.932-27.934i &-39.693-14.646i \\
  $a_{6,I}^{u}(10^{-4})$
      &-558.53-169.85i &-416.46-143.94i &-327.55-124.53i \\
  $a_{6,I}^{c}(10^{-4})$
      &-600.85-40.033i &-448.06-46.988i & -352.8- 47.08i \\
  $a_{6,I}^{u}r_{\chi}(10^{-4})$
      & -484.0-147.19i &-483.15-166.99i &-474.92-180.56i \\
  $a_{6,I}^{c}r_{\chi}(10^{-4})$
      &-520.68-34.691i &-519.81-54.512i &-511.53-68.263i \\ \hline
  $a_{7,I}(10^{-4})$
      & 0.539+0.175i & 1.101+0.086i & 2.419+0.048i \\
  $a_{8,I}^{u}(10^{-4})$
      & 7.383-0.330i & 4.759-0.803i & 2.934-1.165i \\
  $a_{8,I}^{c}(10^{-4})$
      & 7.344-0.209i & 4.632-0.415i & 2.741-0.574i \\
  $a_{8,I}^{u}r_{\chi}(10^{-4})$
      & 6.398-0.286i & 5.521-0.932i & 4.254-1.689i \\
  $a_{8,I}^{c}r_{\chi}(10^{-4})$
      & 6.364-0.181i & 5.374-0.481i & 3.974-0.832i \\
  $a_{9,I}(10^{-4})$
      &-94.827+0.160i &-91.903+0.092i & -89.29+0.057i  \\
  $a_{10,I}^{u}(10^{-4})$
      & -2.426+0.318i & -9.989-0.309i &-15.972-0.812i  \\
  $a_{10,I}^{c}(10^{-4})$
      & -2.497+0.438i &-10.214+0.073i &-16.314-0.230i
  \end{tabular}
  \end{table}

 \begin{table}[htb]
 \caption{Summary of experimental data for the branching ratios (in units
          of $10^{-6}$) for several charmless hadronic $B{\to}PV$. 
          Inequality denotes $90\%$ C.L. upper limits. The last column is
          for the averages of the uncorrelated measurements}
 \label{tab3}
 \begin{tabular}{lcccc}
     \multicolumn{1}{c}{decay decay}
   & \multicolumn{1}{c}{BaBar} 
   & \multicolumn{1}{c}{Belle}
   & \multicolumn{1}{c}{CLEO }
   & \multicolumn{1}{c}{Averaged} \\ \hline
     ${\overline{B}}^{0}{\to}{\pi}^{\mp}{\rho}^{\pm}$  
         & $28.9{\pm}5.4{\pm}{4.3}$     \cite{0107058}
         & $20.2^{+8.3}_{-6.6}{\pm}3.3$ \cite{0104041}
         & $27.6^{+8.4}_{-7.4}{\pm}4.2$ \cite{0006008}
         & $ 25.9{\pm}4.5$              \\
         & & $(<35.7)$ & & \\
     $B^{-}{\to}{\pi}^{-}{\rho}^{0}$  
         & $<39$                        \cite{0105019}
         & $11.2^{+5.3}_{-4.8}{\pm}1.9$ \cite{0104041}
         & $10.4^{+3.3}_{-3.4}{\pm}2.1$ \cite{0006008}
         & $10.7{\pm}3.2$               \\
         & & $(<28.8)$ & & \\
     $B^{-}{\to}{\pi}^{-}{\omega}$
         & $6.6^{+2.1}_{-1.8}{\pm}0.7$  \cite{0108017}
         & $<9.4$                       \cite{0104041}
         & $11.3^{+3.3}_{-2.9}{\pm}1.4$ \cite{0006008}
         & $8.0{\pm}1.7$                \\
     ${\overline{B}}^{0}{\to}K^{-}{\rho}^{+}$
         &
         & $15.8^{+5.1+1.7}_{-4.6-3.0}$ \cite{0115}
         & $16.0^{+7.6}_{-6.4}{\pm}2.8$ \cite{0006008}
         & $15.9{\pm}4.4$               \\
         & & & $(<32)$ & \\
     ${\overline{B}}^{0}{\to}{\overline{K}}^{0}{\omega}$
         & $6.4^{+3.6}_{-2.8}{\pm}0.8$  \cite{0108017} 
         & 
         & $10.0^{+5.4}_{-4.2}{\pm}1.4$ \cite{0006008}
         & $7.5{\pm}2.8$                \\
         & $(<13)$ & & $(<21)$ & \\
     ${\overline{B}}^{0}{\to}{\overline{K}}^{0}{\phi}$
         & $8.1^{+3.1}_{-2.5}{\pm}0.8$ \cite{0105001} 
         & $8.9^{+3.4}_{-2.7}{\pm}1.0$ \cite{0111037}
         & $5.4^{+3.7}_{-2.7}{\pm}0.7$ \cite{0101032}
         & $7.5{\pm}1.8$               \\
         & & & $(<12.3)$ & \\
     ${\overline{B}}^{0}{\to}{\eta}{\overline{K}}^{{\ast}0}$ 
         & $19.8^{+6.5}_{-5.6}{\pm}1.7$ \cite{0107037}
         & $21.2^{+5.4}_{-4.7}{\pm}2.0$ \cite{0111037}
         & $13.8^{+5.5}_{-4.6}{\pm}1.6$ \cite{9912059}
         & $18.0{\pm}3.2$               \\
     $B^{-}{\to}{\pi}^{-}{\overline{K}}^{{\ast}0}$  
         & $15.5{\pm}3.4{\pm}1.8$      \cite{0109007} 
         & $19.4^{+4.2+2.1+3.5}_{-3.9-2.1-6.8}$ \cite{0201007}
         & $7.6^{+3.5}_{-3.0}{\pm}1.6$ \cite{0006008}
         & $14.3{\pm}2.4$              \\
         & & & $(<16)$ & \\
     $B^{-}{\to}{\eta}K^{{\ast}-}$  
         & $22.1^{+11.1}_{-9.2}{\pm}3.3$ \cite{0107037} 
         & 
         & $26.4^{+9.6}_{-8.2}{\pm}3.3$ \cite{9912059}
         & $24.5{\pm}7.1$                \\
         & $(<33.9)$ & & & \\
     $B^{-}{\to}K^{-}{\rho}^{0}$
         & $<29$                       \cite{0105019} 
         & $<13.5$                     \cite{0104041}
         & $8.4^{+4.0}_{-3.4}{\pm}1.8$ \cite{0006008}
         & $8.4{\pm}4.1$               \\
         & & & $(<17)$ &\\
     $B^{-}{\to}K^{-}{\omega}$
         & $1.4^{+1.3}_{-1.0}{\pm}0.3$ \cite{0108017}
         & $<10.5$                     \cite{0104041}
         & $3.2^{+2.4}_{-1.9}{\pm}0.8$ \cite{0006008}
         & $1.8{\pm}1.1$               \\
         & $(<4)$ & & $(<7.9)$ & \\
     $B^{-}{\to}K^{-}{\phi}$
         & $7.7^{+1.6}_{-1.4}{\pm}0.8$  \cite{0105001}
         & $11.2^{+2.2}_{-2.0}{\pm}1.4$ \cite{0111037}
         & $5.5^{+2.1}_{-1.8}{\pm}0.6$  \cite{0101032}
         & $7.9{\pm}1.2$                \\
     ${\overline{B}}^{0}{\to}{\pi}^{+}K^{{\ast}-}$
         &
         & $26.0{\pm}8.3{\pm}3.5$ \cite{0115}
         & $22^{+8+4}_{-6-5}$     \cite{9908018}
         & $23.8{\pm}6.1$         \\
 \end{tabular}
 \end{table}

 \begin{table}
 \caption{CP-averaged branching ratios (in units of $10^{-6}$) of decays
          $B{\to}PV$ for $b{\to}d$ transitions with central values of
          various parameters. The results in columns $2{\sim}4$ are
          calculated with $A=0.819$, ${\lambda}=0.2237$, 
          $\bar{\rho}=0.218$, and $\bar{\eta}=0.316$, while the results
          in columes $5{\sim}7$ are computed with $A=0.83$, 
          ${\lambda}=0.222$, $\bar{\rho}=0.05$, and $\bar{\eta}=0.381$.}
 \label{tab4}
 \begin{tabular}{lccccccc}
     \multicolumn{1}{c}{decay}
   & NF & \multicolumn{2}{c}{QCDF}
   & NF & \multicolumn{2}{c}{QCDF}
   & \\ \cline{2-2}\cline{3-4}\cline{5-5}\cline{6-7}
   \multicolumn{1}{c}{modes}
   & ${\cal B}r$ & ${{\cal B}r}^{f}$ & ${{\cal B}r}^{f+a}$
   & ${\cal B}r$ & ${{\cal B}r}^{f}$ & ${{\cal B}r}^{f+a}$
   & \multicolumn{1}{c}{Exp.} \\ \hline
     ${\overline{B}}^{0}{\to}K^{0}{\overline{K}}^{{\ast}0}$
          & 0.023 & 0.034 & 0.040
          & 0.034 & 0.050 & 0.061
          & --- \\
     ${\overline{B}}^{0}{\to}{\overline{K}}^{0}K^{{\ast}0}$
          & 0.132 & 0.198 & 0.210
          & 0.191 & 0.270 & 0.293
          & --- \\
     ${\overline{B}}^{0}{\to}K^{\pm}K^{{\ast}{\mp}}$
          & ---   & ---   & 0.019
          & ---   & ---   & 0.019
          & --- \\
     ${\overline{B}}^{0}{\to}{\pi}^{-}{\rho}^{+}$
          & 9.231 & 9.685 & 10.13
          & 9.31  & 9.836 & 10.28
          & see Tab.\ref{tab3} \\
     ${\overline{B}}^{0}{\to}{\pi}^{+}{\rho}^{-}$
          & 21.59 & 22.72 & 23.36
          & 19.8  & 20.70 & 21.34
          & see Tab.\ref{tab3} \\
     ${\overline{B}}^{0}{\to}{\pi}^{0}{\rho}^{0}$
          & 0.452 & 0.132 & 0.139
          & 0.434 & 0.124 & 0.127
          & $<5.5$ \cite{0006008} \\
     ${\overline{B}}^{0}{\to}{\pi}^{0}{\omega}$
          & 0.022 & 0.035 & 0.022
          & 0.016 & 0.027 & 0.027
          & $<3$ \cite{0108017} \\
     ${\overline{B}}^{0}{\to}{\pi}^{0}{\phi}$
          & 0.0003 & 0.0006 & ---
          & 0.0004 & 0.0008 & ---
          & $<5$ \cite{PDG2000} \\
     ${\overline{B}}^{0}{\to}{\eta}{\rho}^{0}$
          & 0.002 & 0.010 & 0.037
          & 0.004 & 0.014 & 0.042
          & $<10$ \cite{PDG2000} \\
     ${\overline{B}}^{0}{\to}{\eta}^{\prime}{\rho}^{0}$
          & 0.009 & 0.019 & 0.040
          & 0.007 & 0.016 & 0.030
          & $<12$ \cite{PDG2000} \\
     ${\overline{B}}^{0}{\to}{\eta}{\omega}$
          & 0.253 & 0.113 & 0.129
          & 0.206 & 0.092 & 0.100
          & $<12$ \cite{PDG2000} \\
     ${\overline{B}}^{0}{\to}{\eta}^{\prime}{\omega}$
          & 0.161 & 0.061 & 0.068
          & 0.162 & 0.063 & 0.072
          & $<60$ \cite{PDG2000} \\
     ${\overline{B}}^{0}{\to}{\eta}{\phi}$
          & 0.0001 & 0.0004 & 0.0005
          & 0.0002 & 0.0006 & 0.0007
          & $<9$  \cite{PDG2000} \\
     ${\overline{B}}^{0}{\to}{\eta}^{\prime}{\phi}$
          & 0.0001 & 0.0003 & 0.0002
          & 0.0001 & 0.0004 & 0.0003
          & $<31$ \cite{PDG2000} \\
     $B^{-}{\to}{\pi}^{-}{\rho}^{0}$
          & 7.758 & 6.464 & 6.498
          & 8.125 & 6.949 & 6.954
          & see Tab.\ref{tab3} \\ 
     $B^{-}{\to}{\pi}^{-}{\omega}$
          & 7.988 & 7.08 & 6.977
          & 7.292 & 6.349 & 6.260
          & see Tab.\ref{tab3} \\
     $B^{-}{\to}{\pi}^{-}{\phi}$
          & 0.0006 & 0.0012 & ---
          & 0.0009 & 0.0017 & ---
          & $<5$ \cite{PDG2000} \\
     $B^{-}{\to}{\pi}^{0}{\rho}^{-}$
          & 14.64 & 13.55 & 13.54
          & 13.05 & 11.91 & 11.97
          & $<43$ \cite{0006008} \\ 
     $B^{-}{\to}{\eta}   {\rho}^{-}$
          & 6.627 & 5.879 & 5.798
          & 6.197 & 5.489 & 5.415
          & $<15$ \cite{9912059} \\ 
     $B^{-}{\to}{\eta}^{\prime}{\rho}^{-}$
          & 4.954 & 4.424 & 4.366
          & 4.935 & 4.494 & 4.438
          & $<33$ \cite{9912059} \\
     $B^{-}{\to}K^{0}K^{{\ast}-}$
          & 0.025 & 0.036 & 0.045
          & 0.036 & 0.054 & 0.061
          & --- \\ 
     $B^{-}{\to}K^{-}K^{{\ast}0}$
          & 0.141 & 0.20  & 0.211
          & 0.204 & 0.274 & 0.311
          & $<5.3$ \cite{0006008}
 \end{tabular}
 \end{table}

 \begin{table}
 \caption{CP-averaged branching ratios (in units of $10^{-6}$) of decays
          $B{\to}PV$ for $b{\to}s$ transitions with central values of
          various parameters. The results in columns $2{\sim}4$ are
          calculated with $A=0.819$, ${\lambda}=0.2237$,
          $\bar{\rho}=0.218$, and $\bar{\eta}=0.316$, while the results
          in columns $5{\sim}7$ are computed with $A=0.83$,
          ${\lambda}=0.222$, $\bar{\rho}=0.05$, and $\bar{\eta}=0.381$.}
 \label{tab5}
 \begin{tabular}{lccccccc}
     \multicolumn{1}{c}{decay}
   & NF & \multicolumn{2}{c}{QCDF}
   & NF & \multicolumn{2}{c}{QCDF}
   & \\ \cline{2-2}\cline{3-4}\cline{5-5}\cline{6-7}
   \multicolumn{1}{c}{modes}
   & ${\cal B}r$ & ${{\cal B}r}^{f}$ & ${{\cal B}r}^{f+a}$
   & ${\cal B}r$ & ${{\cal B}r}^{f}$ & ${{\cal B}r}^{f+a}$
   & \multicolumn{1}{c}{Exp.} \\ \hline
     ${\overline{B}}^{0}{\to}K^{-}{\rho}^{+}$
          & 1.485 & 1.848 & 2.008
          & 1.081 & 1.321 & 1.501
          & see Tab.\ref{tab3} \\
     ${\overline{B}}^{0}{\to}{\overline{K}}^{0}{\rho}^{0}$
          & 0.918 & 1.184 & 1.256
          & 1.038 & 1.239 & 1.297
          & $<39$ \cite{PDG2000} \\
     ${\overline{B}}^{0}{\to}{\overline{K}}^{0}{\omega}$
          & 0.034 & 0.083 & 0.007
          & 0.026 & 0.076 & 0.012
          & see Tab.\ref{tab3} \\
     ${\overline{B}}^{0}{\to}{\overline{K}}^{0}{\phi}$
          & 3.663 & 5.945 & 6.703
          & 3.589 & 5.833 & 6.569
          & see Tab.\ref{tab3} \\
     ${\overline{B}}^{0}{\to}{\pi}^{+}K^{{\ast}-}$
          & 1.838 & 2.411 & 2.743
          & 3.281 & 4.077 & 4.36
          & see Tab.\ref{tab3} \\
     ${\overline{B}}^{0}{\to}{\pi}^{0}{\overline{K}}^{{\ast}0}$
          & 0.533 & 0.744 & 0.896
          & 0.459 & 0.714 & 0.875
          & $<3.6$ \cite{0006008} \\
     ${\overline{B}}^{0}{\to}{\eta}{\overline{K}}^{{\ast}0}$
          & 2.072 & 2.681 & 2.972
          & 2.15  & 2.67  & 2.927
          & see Tab.\ref{tab3} \\
     ${\overline{B}}^{0}{\to}{\eta}^{\prime}{\overline{K}}^{{\ast}0}$
          & 0.759 & 1.717 & 1.891
          & 0.689 & 1.662 & 1.84
          & $<24$ \cite{9912059} \\
     $B^{-}{\to}{\pi}^{-}{\overline{K}}^{{\ast}0}$
          & 2.583 & 3.497 & 3.814
          & 2.531 & 3.433 & 3.731
          & see Tab.\ref{tab3} \\
     $B^{-}{\to}{\pi}^{0}K^{{\ast}-}$
          & 1.852 & 2.317 & 2.489
          & 3.067 & 3.543 & 3.667
          & $<31$\cite{0006008} \\
     $B^{-}{\to}{\eta}K^{{\ast}-}$
          & 1.777 & 2.247 & 2.591
          & 2.564 & 3.022 & 3.306
          & see Tab.\ref{tab3} \\
     $B^{-}{\to}{\eta}^{\prime}K^{{\ast}-}$
          & 1.446 & 2.664 & 2.83
          & 1.046 & 2.091 & 2.277
          & $<35$ \cite{9912059} \\
     $B^{-}{\to}{\overline{K}}^{0}{\rho}^{-}$
          & 0.403 & 0.598 & 0.789
          & 0.395 & 0.585 & 0.777
          & $<48$ \cite{PDG2000} \\
     $B^{-}{\to}K^{-}{\rho}^{0}$
          & 0.453 & 0.426 & 0.528
          & 0.609 & 0.503 & 0.631
          & see Tab.\ref{tab3} \\
     $B^{-}{\to}K^{-}{\omega}$
          & 0.583 & 0.530 & 0.435
          & 0.580 & 0.565 & 0.495
          & see Tab.\ref{tab3} \\
     $B^{-}{\to}K^{-}{\phi}$
          & 3.911 & 6.346 & 7.179
          & 3.831 & 6.227 & 7.02
          & see Tab.\ref{tab3}
 \end{tabular}
 \end{table}

 \begin{table}
 \caption{CP-violating asymmetry parameters $a_{{\epsilon}^{\prime}}$,
          and $a_{{\epsilon}+{\epsilon}^{\prime}}$ (in units of $10^{-2}$)
          for decays ${\overline{B}}^{0}{\to}PV$ with central values of
          various parameters within the  QCDF framework. The results in
          columns $2{\sim}5$ are calculated with $A=0.819$,
          ${\lambda}=0.2237$, $\bar{\rho}=0.218$, and $\bar{\eta}=0.316$,
          while the results in columns $6{\sim}9$ are calculated with
          $A=0.83$, ${\lambda}=0.222$, $\bar{\rho}=0.05$, and
          $\bar{\eta}=0.381$.}
 \label{tab6}
 \begin{tabular}{ccccccccc}
     \multicolumn{1}{c}{modes}
   & $a_{{\epsilon}^{\prime}}^{f}$
   & $a_{{\epsilon}^{\prime}}^{f+a}$
   & $a_{{\epsilon}+{\epsilon}^{\prime}}^{f}$
   & $a_{{\epsilon}+{\epsilon}^{\prime}}^{f+a}$
   & $a_{{\epsilon}^{\prime}}^{f}$
   & $a_{{\epsilon}^{\prime}}^{f+a}$
   & $a_{{\epsilon}+{\epsilon}^{\prime}}^{f}$
   & $a_{{\epsilon}+{\epsilon}^{\prime}}^{f+a}$ \\ \hline
     ${\pi}^{0}{\rho}^{0}$
          & -11.11 &  -6.90 &  48.29 &  51.66
          & -14.01 &  -8.88 & -39.15 & -34.80 \\
     ${\pi}^{0}{\omega}$
          &  19.08 &  80.95 &  94.80 &  32.56
          &  28.82 &  80.02 &  95.35 &  10.70 \\
     ${\eta}{\rho}^{0}$
          &  31.78 & 16.77 & -17.31 &  10.53
          &  28.11 & 17.11 & -87.88 & -75.06 \\
     ${\eta}^{\prime}{\rho}^{0}$
          & -28.02 & -34.78 &  89.23 & 92.18
          & -39.96 & -54.92 &  88.09 & 66.99 \\
     ${\eta}{\omega}$
          &  44.82 &  29.66 &  74.53 &  81.05
          &  64.64 &  45.06 &  16.02 &  26.05 \\
     ${\eta}^{\prime}{\omega}$
          & -20.57 & -16.88 &  32.27 &  26.95
          & -23.60 & -18.81 & -56.70 & -62.16 \\
     $K_{S}^{0}{\rho}^{0}$
          & -5.56  &  -8.92 &  64.39 &  66.25
          & -6.27  & -10.20 &  62.38 &  64.40 \\
     $K_{S}^{0}{\omega}$
          & -41.77 &  86.90 &  79.28 & -48.78
          & -54.10 &  62.18 &  73.15 & -61.69 \\
     $K_{S}^{0}{\phi}$
          &  -0.99 &  -0.97 &  73.24 &  73.40
          &  -1.19 &  -1.17 &  72.91 &  73.11 \\
     ${\pi}^{0}{\phi}$
        & 0 & --- & 2.29 & ---
        & 0 & --- & 1.83 & --- \\
     ${\eta}^{(\prime)}{\phi}$
        & 0 & 0 & 2.29 & 2.29
        & 0 & 0 & 1.83 & 1.83
 \end{tabular}
 \end{table}

 \begin{table}
 \caption{CP-violating asymmetries ${\cal A}_{CP}$ ($\%$) for $B{\to}PV$
         ($b{\to}d$ transitions) with central values of various parameters
         within the  QCDF approach. The results in the 2nd and 3rd columns
         are calculated with $A=0.819$, ${\lambda}=0.2237$,
         $\bar{\rho}=0.218$, and $\bar{\eta}=0.316$, while the results in
         the 4th and 5th columns are computed with $A=0.83$,
         ${\lambda}=0.222$, $\bar{\rho}=0.05$, and $\bar{\eta}=0.381$.}
 \label{tab7}
 \begin{tabular}{lcccc}
     \multicolumn{1}{c}{modes}
   & ${\cal A}_{CP}^{f}$ & ${\cal A}_{CP}^{f+a}$
   & ${\cal A}_{CP}^{f}$ & ${\cal A}_{CP}^{f+a}$ \\ \hline
     $B^{0}{\to}K^{0}_{S}{\overline{K}}^{{\ast}0}$
          & 22.38 & 23.89
          & 19.60 & 20.37 \\
     $B^{0}{\to}K^{0}_{S}K^{{\ast}0}$
          & -11.34 & -11.97
          &  -9.14 &  -9.36 \\
     $B^{0}{\to}K^{\pm}K^{{\ast}{\mp}}$
          & --- &  17.41
          & --- & -25.49 \\
     $B^{0}{\to}{\pi}^{-}{\rho}^{+}$
          & 17.09 & 20.16
          &-13.47 & -9.60 \\
     $B^{0}{\to}{\pi}^{+}{\rho}^{-}$
          & 25.64 & 21.51
          &-26.91 &-31.85 \\
     ${\overline{B}}^{0}{\to}{\pi}^{0}{\rho}^{0}$
          & 15.75 & 20.10
          &-27.78 &-22.36 \\
     ${\overline{B}}^{0}{\to}{\pi}^{0}{\omega}$
          & 57.59 & 68.32
          & 64.21 & 57.29 \\
     ${\overline{B}}^{0}{\to}{\eta}{\rho}^{0}$
          & 12.49 & 15.95
          &-23.52 &-24.58 \\
     ${\overline{B}}^{0}{\to}{\eta}^{\prime}{\rho}^{0}$
          & 24.21 & 21.21
          & 15.88 & -3.92 \\
     ${\overline{B}}^{0}{\to}{\eta}{\omega}$
          & 64.74 & 57.95
          & 49.80 & 41.81 \\
     ${\overline{B}}^{0}{\to}{\eta}^{\prime}{\omega}$
          &  1.95 &  1.83
          &-42.40 &-41.88 \\
     ${\overline{B}}^{0}{\to}{\pi}^{0}{\phi}$
          & 1.09  &  ---
          & 0.87  &  --- \\
     ${\overline{B}}^{0}{\to}{\eta}^{(\prime)}{\phi}$
          & 1.09  & 1.09
          & 0.87  & 0.87 \\
     $B^{-}{\to}{\pi}^{-}{\rho}^{0}$
          &  3.26 & 12.47
          &  3.58 & 13.76 \\
     $B^{-}{\to}{\pi}^{-}{\omega}$
          &  -6.22 &  -6.64
          &  -8.18 &  -8.73 \\
     $B^{-}{\to}{\pi}^{-}{\phi}$
          & 0      & ---
          & 0      & --- \\
     $B^{-}{\to}{\pi}^{0}{\rho}^{-}$
          &  -2.98 &  -9.09
          &  -4.0  & -12.14 \\
     $B^{-}{\to}{\eta}   {\rho}^{-}$
          &  -1.75 &  -2.05
          &  -2.21 &  -2.59 \\
     $B^{-}{\to}{\eta}^{\prime}{\rho}^{-}$
          &   5.03 &   4.94
          &   5.85 &   5.74 \\
     $B^{-}{\to}K^{0}_{S}K^{{\ast}-}$
          &  -8.82 &   8.94
          &  -6.94 &   7.83 \\
     $B^{-}{\to}K^{-}K^{{\ast}0}$
          & -25.28 & -33.23
          & -21.78 & -26.58
 \end{tabular}
 \end{table}

 \begin{table}
 \caption{CP-violating asymmetries ${\cal A}_{CP}$ ($\%$) for $B{\to}PV$
         ($b{\to}s$ transitions) with central values of various parameters
         within the QCDF approach. The results in the 2nd and 3rd columns
         are calculated with $A=0.819$, ${\lambda}=0.2237$,
         $\bar{\rho}=0.218$, and $\bar{\eta}=0.316$, while the results in
         the 4th and 5th columns are computed with $A=0.83$,
         ${\lambda}=0.222$, $\bar{\rho}=0.05$, and $\bar{\eta}=0.381$.}
 \label{tab8}
 \begin{tabular}{lcccc}
     \multicolumn{1}{c}{modes}
   & ${\cal A}_{CP}^{f}$ & ${\cal A}_{CP}^{f+a}$
   & ${\cal A}_{CP}^{f}$ & ${\cal A}_{CP}^{f+a}$ \\ \hline
     ${\overline{B}}^{0}{\to}K^{-}{\rho}^{+}$
          &   1.62 & -34.31
          &   2.68 & -54.21 \\
     ${\overline{B}}^{0}{\to}K_{S}^{0}{\rho}^{0}$
          &  27.04 &  25.73
          &  25.62 &  24.01 \\
     ${\overline{B}}^{0}{\to}K_{S}^{0}{\omega}$
          &  10.51 &  33.46
          &  -0.46 &  11.19 \\
     ${\overline{B}}^{0}{\to}K_{S}^{0}{\phi}$
          &  34.23 &  34.32
          &  33.95 &  34.05 \\
     ${\overline{B}}^{0}{\to}{\pi}^{+}K^{{\ast}-}$
          &  20.57 &  47.0
          &  14.36 &  34.92 \\
     ${\overline{B}}^{0}{\to}{\pi}^{0}{\overline{K}}^{{\ast}0}$
          &  -9.49 &  -9.51
          & -11.69 & -11.50 \\
     ${\overline{B}}^{0}{\to}{\eta}{\overline{K}}^{{\ast}0}$
          &   5.05 &   5.52
          &   5.99 &   6.61 \\
     ${\overline{B}}^{0}{\to}{\eta}^{\prime}{\overline{K}}^{{\ast}0}$
          &  -5.51 &  -5.56
          &  -6.72 &  -6.75 \\
     $B^{-}{\to}{\pi}^{-}{\overline{K}}^{{\ast}0}$
          &   0.97 &   1.23
          &   1.17 &   1.49 \\
     $B^{-}{\to}{\pi}^{0}K^{{\ast}-}$
          &  18.86 &  35.57
          &  14.56 &  28.51 \\
     $B^{-}{\to}{\eta}K^{{\ast}-}$
          &   7.73 &  29.28
          &   6.79 &  27.09 \\
     $B^{-}{\to}{\eta}^{\prime}K^{{\ast}-}$
          & -13.78 & -20.39
          & -20.73 & -29.92 \\
     $B^{-}{\to}K^{0}_{S}{\rho}^{-}$
          &   0.31 &  -0.34
          &   0.38 &  -0.41 \\
     $B^{-}{\to}K^{-}{\rho}^{0}$
          &   2.88 & -80.25
          &   2.87 & -79.31 \\
     $B^{-}{\to}K^{-}{\omega}$
          &  59.85 & -19.92
          &  66.26 & -20.63 \\
     $B^{-}{\to}K^{-}{\phi}$
          &   0.99 &   1.17
          &   1.19 &   1.41
 \end{tabular}
 \end{table}

 \end{document}